\documentclass[aps,prl,amsmath,amssymb,twocolumn]{revtex4-1}
\usepackage{graphicx}
\begin{document}

%\begin{frontmatter}

\title{Water Dynamics at Rough Interfaces}

\author{M. Rosenstihl}
\author{K. K\"ampf}
\author{F. Klameth}
\author{M. Sattig}
\author{M. Vogel}
\address{Institut f\"ur Festk\"orperphysik, Technische Universit\"at Darmstadt, Hochschulstr.\ 6, 64289 Darmstadt, Germany}

\begin{abstract}
We use molecular dynamics computer simulations and nuclear magnetic resonance experiments to investigate the dynamics of water at interfaces of molecular roughness and low mobility. We find that, when approaching such interfaces, the structural relaxation of water, i.e., the $\alpha$ process, slows down even when specific attractive interactions are absent. This prominent effect is accompanied by a smooth transition from Vogel to Arrhenius temperature dependence and by a growing importance of jump events. Consistently, at protein surfaces, deviations from Arrhenius behavior are weak when free water does not exist. Furthermore, in nanoporous silica, a dynamic crossover of liquid water occurs when a fraction of solid water forms near 225\,K and, hence, the liquid dynamics changes from bulk-like to interface-dominated. At sufficiently low temperatures, water exhibits a quasi-universal $\beta$ process, which is characterized by an activation energy of $E_a\!=\!0.5$\,eV and involves anisotropic reorientation about large angles. As a consequence of its large amplitude, the faster $\beta$ process destroys essentially all orientational correlation, rendering observation of a possible slower $\alpha$ process difficult in standard experiments. Nevertheless, we find indications for the existence of structural relaxation down to a glass transition of interfacial water near 185\,K. Hydrated proteins show a highly restricted backbone motion with an amplitude, which decreases upon cooling and vanishes at comparable temperatures, providing evidence for a high relevance of water rearrangements in the hydration shell for secondary protein relaxations.  
\end{abstract}

%\begin{keyword}
%Confined water, mesoporous silica, hydrated proteins, nuclear magnetic resonance, molecular dynamics simulations
%\end{keyword}

%\end{frontmatter}

%\linenumbers

\maketitle

\section{Introduction}

The interaction of water with interfaces is of utmost importance in nature and technology \cite{Debenedetti03, Bagchi05}. Thereby, the interfaces can be hydrophilic or hydrophobic and they can differ with respect to extension and stiffness. Water may reside near vast membrane and clay surfaces or it may be intimately mixed at various concentrations with small molecules exhibiting comparable or different mobility. Although there is no doubt that the dynamics of water depends on the immediate environment, it is still unclear in which way the interface properties determine the degree and range of possible deviations from the bulk behavior.

All these situations can be found for water-protein mixtures. Proteins feature specific sequences of hydrophilic and hydrophobic amino acids, they differ regarding to molecular size and stiffness, and they are miscible with water at various concentrations, ranging from the dilute regime to the crowded regime. Hence, water-protein mixtures may be considered as ideal systems to gain insights into interactions between water and interfaces. In particular, they yield information about dynamical couplings between water and protein molecules, which were considered crucial for biological functions \cite{Frauenfelder09}.

Aside from great benefits, usage of proteins has some shortcomings for fundamental approaches to the behavior of water at interfaces. Specifically, the variation of amino acids associated with the primary structure and the packing of protein molecules in a disordered material result in a diversity of local environments and, hence, in overlapping effects. Consequently, complementary studies on materials with tailored homogeneous environments are advisable. In this context, it proved useful to mix water with appropriate small molecules or to embed it in mesoporous silica, which exhibit cylindrical confinements with defined and tunable diameters \cite{Angell02, Angell08, Ngai13}.            

An understanding of water-interface interactions is also important to decide to which extent studies for confinements and mixtures provide insights into properties of liquid water in the bulk, in particular, in the no-man's land, 150--235\,K, where crystallization can be avoided for confined and mixed water, but not for the bulk liquid. Access to the behavior of liquid water in the no-man's land is of major importance since the water anomalies were traced back to the existence of a second critical point, which is associated with a liquid-liquid (LL) phase transition between high-density and low-density forms of water in the deeply supercooled regime, most probably near 225\,K \cite{Poole92, Mishima98}. Such LL phase transition should manifest itself in a fragile-to-strong (FS) dynamic crossover, which was postulated in view of different water behaviors above and below the no man's land \cite{Ito99, Angell08}. 

Previous investigations of water in confinements and mixtures did not provide a uniform picture. Some workers argued that confined and mixed waters exhibit a LL transition and a FS crossover near 225\,K, which are general phenomena and, hence, relevant for the bulk liquid \cite{Stanley10, Mallamace12, Zanotti05, Liu05, Chen06, Yoshida08, Kumar06, Gallo10}, while other workers challenged this conjecture \cite{Cerveny04, Swenson06, Hedstrom07, Cerveny08, Khodadadi08, Pawlus08, Bruni11, Gainaru09, Lusceac11, Vogel08, Sattig14, Doster10, Ngai10, Capaccioli11}. In detail, neutron scattering (NS) approaches to water in silica and protein confinements reported sharp kinks in temperature-dependent correlation times and molecular displacements and took them as evidence for a FS crossover related to a LL transition \cite{Zanotti05, Liu05, Chen06, Yoshida08}. This opinion received support from molecular dynamics (MD) simulations for such systems \cite{Kumar06, Gallo10}. By contrast, NS results \cite{Doster10} did not show a FS crossover when analyzing the experimental data in a different way and MD work \cite{Limmer11} argued that computational results attributed to a LL transition in fact reflect crystallization. Moreover, dielectric spectroscopy (DS) studies on water in various environments did not observe a sharp kink in the temperature dependence at 225\,K, but, if at all, a mild bending at variable temperatures \cite{Cerveny04, Swenson06, Hedstrom07, Cerveny08, Khodadadi08, Pawlus08, Bruni11, Gainaru09, Lusceac11}. Specifically, these approaches found an Arrhenius law with a common activation energy of $E_a\!=0.5\,$eV at sufficiently low temperatures. While this behavior spreads over the whole temperature range for some systems, it gives way to fragile behavior when increasing the temperature for other systems. The latter change of the temperature dependence was attributed to a passage from a bulk-like to a confinement-affected $\alpha$ process \cite{Cerveny08}, to a crossover from the $\alpha$ process to the $\beta$ process \cite{Swenson06, Pawlus08} or to changes of the $\beta$ process at the glass transition temperature $T_g$ \cite{Ngai10, Capaccioli11}.  

An inconsistent picture of water behavior also emerged from nuclear magnetic resonance (NMR) investigations. $^1$H NMR works reported evidence for changes in structural and dynamical properties of confined water in response to a LL transition \cite{Mallamace12, Mallamace08}. In particular, diffusion coefficients obtained from measurements in magnetic gradient fields were found to exhibit a kink in the temperature dependence near 225\,K \cite{Mallamace06, Mallamace07}. Our $^1$H NMR diffusion study, however, observed that spin relaxation effects interfere with a reliable determination of diffusion coefficients below 225\,K \cite{Rosenstihl11}. Also, in $^2$H NMR studies, some results were interpreted in terms of a LL transition \cite{Hwang07}, while other findings provided evidence against the existence of this phenomenon for water in mesoporous silica and protein matrices \cite{Lusceac11, Vogel08, Sattig14}.           

To improve our knowledge about interactions of water with interfaces, it is advisable to characterize not only rates, but also mechanisms for water dynamics in various environments. In particular, for an interpretation of changes in temperature-dependent correlation times in terms of a FS crossover, it is an essential precondition that the $\alpha$ process is probed above and below the crossover temperature. Here, we exploit the fact that MD simulations and NMR experiments enable detailed insights into both time constants and motional mechanisms of molecular dynamics. We use these capabilities to gain information about dynamics of water (H2O and D2O) in hydration shells of proteins, explicitly, at collagen (COL), elastin (ELA), and myoglobin (MYO) surfaces, in nanoscopic confinements, in particular, in mesoporous silica (MCM-41) with various pore diameters, and, to some extent, in mixtures with small molecules of low mobility. 

In MD simulations, we utilize that full microscopic information is available for tailored model systems to determine the character and range of interface effects for water in various environments. Due to the available computer performance, these studies are nowadays restricted to the weakly supercooled regime with correlation times $\tau\!<\!100\,$ns. In NMR work, we exploit that $^1$H NMR in gradient fields and $^2$H NMR in homogeneous fields provide access to water diffusion on mesoscopic scales and water reorientation on local scales, respectively. 

\section{Theoretical Background}\label{Theory}

Water rotational motion can be characterized based on the autocorrelation functions
\begin{equation}
F_l(t)=\frac{\langle P_l[\cos\theta(0)] P_l[\cos\theta(t)]\rangle}{\langle P_l[\cos\theta(0)] P_l[\cos\theta(0)]\rangle}
\end{equation}
Here, $P_l$ is the Legendre polynomial of rank $l$, $\theta$ is an angle characterizing the molecular orientation, and the pointed brackets indicate an ensemble average throughout this article. We denote the corresponding rotational correlation times as $\tau_l$. Furthermore, the associated spectral densities and dynamic susceptibilities are referred to as $J_l(\omega)$ and $\chi_l(\omega)$, respectively. While $F_1(t)$ is accessible from DS studies, $F_2(t)$ is available from NMR approaches. In MD simulations, knowledge of the atomic trajectories allows us to calculate all these quantities. 

In our $^2$H NMR studies, we observe deuterons in heavy water and detect their quadrupolar frequency $\omega_Q$ associated with the interaction between the electric quadrupole moment of the nucleus and an electric field gradient at the nuclear site. This frequency is approximately given by \cite{Schmidt-Rohr94}
\begin{equation} \label{om}
  \omega_Q=\pm\frac{\delta}{2}(3\cos^2\theta-1) \propto P_2(\cos \theta)
\end{equation}
where $\delta$ describes the strength of the quadrupolar interaction and $\theta$ characterizes the orientation of the water molecule, explicitly, the angle between the O--D bond and the applied $\mathbf{B}_0$ field. Thus, the fluctuations of $\omega_Q$ reflect the reorientation of the water molecule. $^2$H spin-lattice relaxation (SLR) times $T_1$ are determined by the spectral density of isotropic reorientation according to 
\begin{equation}\label{T1}
\frac{1}{T_{1}}=\frac{2}{15}\,\delta^2\left[J_2(\omega_0)+4 J_2(2\omega_0)\right]
\end{equation}
with the Larmor frequency $\omega_0$. $T_1$ is at a minimum when the rotational correlation time $\tau_2$ obeys $\omega_0\tau_2\!\approx\!1$. $^2$H stimulated-echo (STE) experiments enable direct measurements of the rotational correlation functions \cite{Schmidt-Rohr94, Fleischer94, Boehmer01}
\begin{eqnarray}
  F_2^{cc}(t_m,t_p)&\propto&\langle\,\cos\left[\,\omega_Q(0)t_p\right]\cos\left[\, \omega_Q(t_m)t_p\right]\rangle  \label{cc}
  \\
  F_2^{ss}(t_m,t_p)&\propto&\langle\,\sin\left[\,\omega_Q(0)t_p\right]\sin\left[\, \omega_Q(t_m)t_p\right]\rangle
  \label{ss}
\end{eqnarray}
In STE experiments, changes of $\omega_Q$ due to water reorientation are probed by a variation of the mixing time $t_m$ for a fixed length of the evolution time $t_p$, which determines the angular resolution of the experiment \cite{Fleischer94, Boehmer01}, in analogy with the momentum transfer $Q$ in scattering studies \cite{Fujara86}. The correlation function $F_2(t_m)$ is obtained from $F_2^{ss}(t_m)$ in the limit $t_p\!\rightarrow\!0$ where $\sin(\omega_Q t_p)\!\propto\! P_2(\cos \theta)$.

In $^1$H NMR, we apply static magnetic field gradients (FG). Then, the observed frequency is not governed by the molecular orientation, but by the molecular position within the gradient field \cite{Callaghan91}. As a consequence, STE experiments probe frequency changes resulting from the translational diffusion of the water molecule and, hence, provide access to the self-diffusion coefficient $D$.   
   
\section{Results}
\subsection{Computational Studies of Water Dynamics}\label{Simulation}

First, we exploit that MD simulations enable studies of water in tailored environments. We compare results for water at elastin surfaces and in silica pores with findings for water in 'neutral' matrices. The latter matrices are available in MD simulations when using equilibrium configurations of bulk water and pinning suitable subsets of water molecules \cite{Klameth13, Klameth14}. Thereby, we choose the pinned molecules so as to obtain cylindrical pores (CP) or random matrices (RM). Thus, the interactions and geometries imprinted by the confining matrices on the confined water differ in our studies. While specific interactions of water with protein and silica surfaces usually lead to distortions of water structure \cite{Smolin05, Gallo10}, e.g., to density oscillations in the direction perpendicular to the surface, neutral confinements do not cause such structural changes \cite{Klameth13, Klameth14}. In the following, we ascertain effects of these interfaces on water dynamics, in particular, on the rotational motion, which is probed in the later NMR approaches, while the translational motion was elucidated in our previous studies \cite{Klameth13, Klameth14}. To obtain detailed insights, we perform spatially resolved analyses, i.e., we distinguish between water molecules at various distances $d$ to the nearest interface. While the SPC model \cite{SPC} of water is used in a mixture with an ELA-like peptide \cite{Vogel09}, the SPC/E model \cite{SPCE} is employed in the other systems. 

\begin{figure} 
\includegraphics[width=8.7cm]{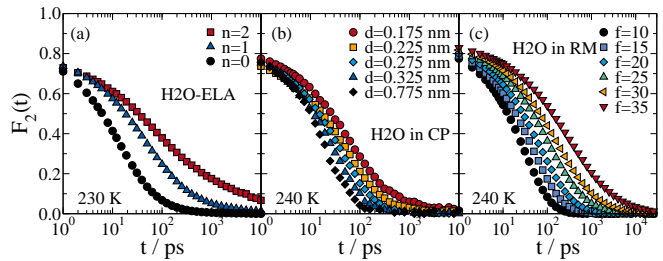}
\centering
\caption{Correlation functions $F_2(t)$ of water molecules residing at various distances from an interface at $t\!=\!0$. (a) Mixture of water and elastin ($h\!=\!0.3$\,g/g): Water molecules forming different numbers $n$ of hydrogen bonds with a protein molecule are distinguished. (b) Water in a cylindrical pore (5\,nm diameter) formed by pinned water molecules: Water molecules at various distances $d$ to the nearest pinned water molecule are discriminated. (c) Water in a random matrix formed by pinned water molecules: Various fractions $f$ (in percent) of pinned molecules are considered.}
\label{Fig_1}
\end{figure}

In Fig.\ \ref{Fig_1}, we compare rotational correlation functions $F_2(t)$ of water molecules located at various distances from an interface. In Fig.\ \ref{Fig_1}(a), we see for a H2O-ELA mixture \cite{Vogel09} that the rotational motion slows down when water molecules form hydrogen bonds with a protein molecule and, hence, reside at a water-protein interface. However, specific guest-host attractions are no precondition for a slowdown at interfaces, as becomes clear from studies of water dynamics in neutral confinements. In Fig.\ \ref{Fig_1}(b), it is evident for H2O in CP that $F_2(t)$ decays significantly slower for water in the vicinity of the wall than for that in the center of the pore. Thus, the time scale of water reorientation strongly varies across the pore. In Fig.\ \ref{Fig_1}(c), we consider H2O in RM, mimicking situations when the more mobile component of a mixture moves in a glassy matrix formed by the less mobile component. We observe that water dynamics becomes more sluggish when the fraction $f$ of randomly pinned molecules increases and, hence, the average distance between an unpinned molecule and the nearest pinned molecule decreases. These results reveal that there is a slowdown of water dynamics near walls of molecular roughness and low mobility, which is not a consequence of peculiar attractive forces.

\begin{figure} 
\includegraphics[width=8.7cm]{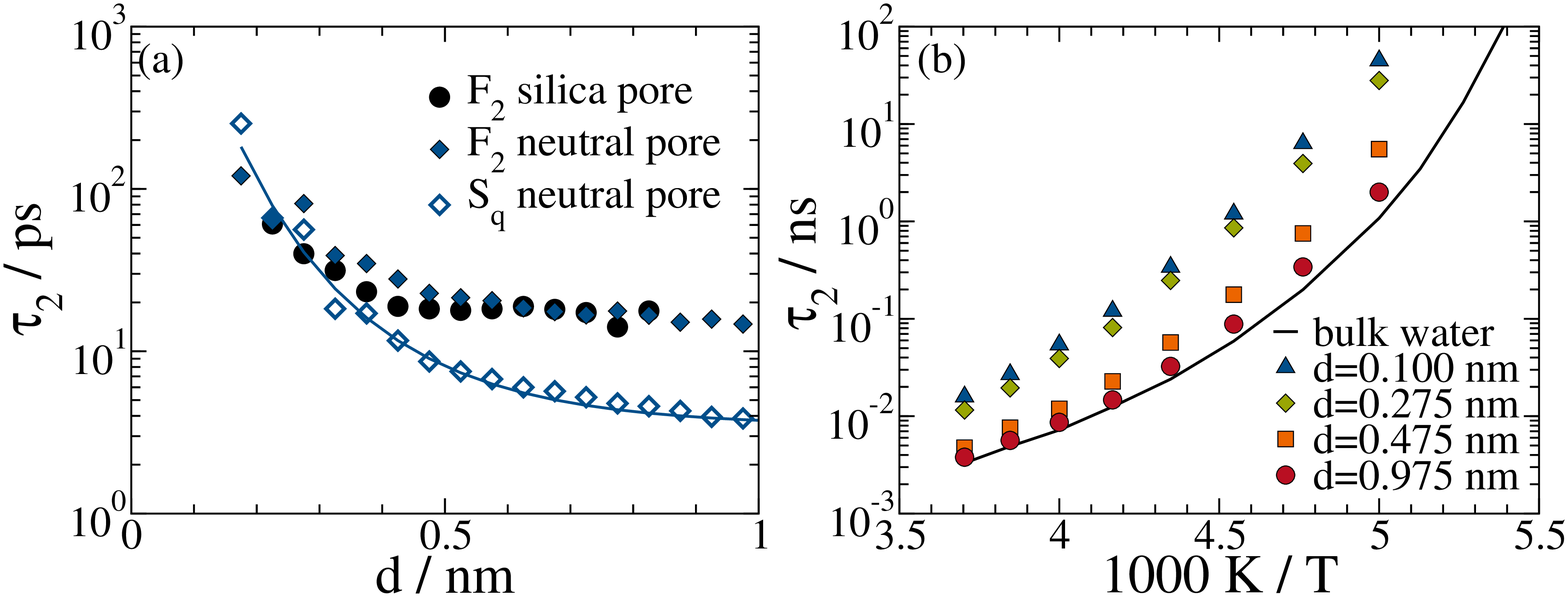}
\caption{Correlation times of water dynamics in different regions of cylindrical confinements: (a) Correlation times $\tau_2$ for the rotational motion as a function of the distance $d$ to the pore wall for water in a silica pore (2\,nm diameter) and in a neutral pore formed by pinned water molecules (5\,nm diameter). For the neutral pore, we also show time constants $\tau$ for the translational motion, which were obtained from incoherent intermediate scattering functions $S_q(t)$ in previous work \cite{Klameth13}. The line is a fit to Eq.\ (\ref{expfit}). (b) Temperature dependence of correlation times $\tau_2$ at different positions across the neutral confinement. The line indicates the bulk behavior. The time constants were determined according to $F_2(t\!=\!\tau_2)\!=\!1/\mathrm{e}$ and $S_q(t\!=\!\tau)\!=\!1/\mathrm{e}$}
\label{Fig_2}
\end{figure} 

For an analysis of the degree and range of this slowdown, we separately determine correlation times of water reorientation in different regions of cylindrical confinements, explicitly, at various distances $d$ to the pore walls. The dynamic profiles $\tau_2(d)$ resulting for water in neutral and silica pores are presented in Fig.\ \ref{Fig_2}(a). It can be seen that the correlation time varies at the pore walls, while it is constant in the pore center. The exact shape of the dynamic profile, however, differs among the confinements and depends on temperature. We observe that the slowdown is stronger and extends to larger distances for the neutral confinement than for the silica confinement. For translational motion, the slowdown at the pore walls is even more prominent \cite{Klameth13}. It can be described by the empirical relation
\begin{equation}
\ln\left (\frac{\tau}{\tau_\infty} \right) \propto \exp \left( - \frac{d}{\xi} \right)  
\label{expfit}
\end{equation}
which was proposed for atomic glass formers \cite{Scheidler04, Berthier12}. In this equation, $\tau_\infty$ characterizes the time scale of water dynamics at large distances and the length scale $\xi$ specifies the range of the wall effect, which was found to nearly double when the temperature is decreased from 270\,K to 200\,K \cite{Klameth13}. For the studied confinements and temperatures, the impact of the walls on water dynamics is very striking for two layers of water at the interfaces, but discernible effects exist up to larger distances of about 1\,nm. Accordingly, bulk behavior is not recovered in the center of very small pores with such diameters \cite{Klameth13}.

In Fig.\ \ref{Fig_2}(b), we observe exemplary for H2O in CP that the slowdown of water dynamics at rough walls is accompanied by a change of the temperature dependence. Specifically, the fragility is high in the pore center, where bulk behavior is recovered, while it is low near the pore walls, where the behavior is closer to an Arrhenius law. Such alteration of the temperature dependence upon approaching an interface is shared by rotational and translation water motion and it occurs at neutral, protein, and silica surfaces \cite{Vogel09, Klameth13, Klameth14, Gallo10}. The effect is accompanied by a change of the mechanism for water motion. Specifically, water dynamics evolves from diffusive motion to jump motion when decreasing the distance to solid surfaces of molecular roughness \cite{Vogel09, Klameth13}. Near interfaces at low temperatures, translational jumps about the intermolecular distance and rotational jumps about the tetrahedral angle prevail, where the latter type of motion includes $\pi$ flips about the molecular symmetry axis. 

The atoms of the studied neutral, protein, and silica surfaces impose a static energy landscape for the water motion in the neighborhood. For example, they provide well defined preferred sites for the formation of hydrogens bonds. Such static energy landscape not only imprints static density correlations near surfaces \cite{Klameth14}, but also decreases the capability for cooperative structural rearrangements. Rather, energy barriers need to be overcome when moving between configurations that are well adapted to the wall structure. Consequently, the dynamics becomes slower, the fragility is reduced, and the mechanism changes from liquid-like diffusion to solid-like hopping, in harmony with our simulation results. In the literature \cite{Drossel14}, a strong influence of wall roughness was observed for water dynamics near hydrophilic surfaces. Specifically, when reducing the roughness of the wall and keeping the average potential energy at a given distance constant, the slowdown of interfacial water becomes significantly weaker. Hence, simple excluded-volume effects or attractive water-matrix interactions are not sufficient to explain the altered dynamical behavior of water at hydrophilic surfaces, consistent with our findings for neutral confinements. Rather, these findings support our conclusion that water dynamics near solid surfaces of molecular roughness is governed by static energy landscapes imposed by the wall atoms. In addition, it is important to take into account that the structure of water is disturbed to different degrees at various surfaces. One may expect that a disturbance of water structure counteracts the slowdown of water dynamics. In agreement this argument, we observed that the mobility of water molecules is more reduced at neutral surfaces, which do not distort the structure, than at silica surfaces, where significant structural changes exist.

\begin{figure} 
\includegraphics[width=8.7cm]{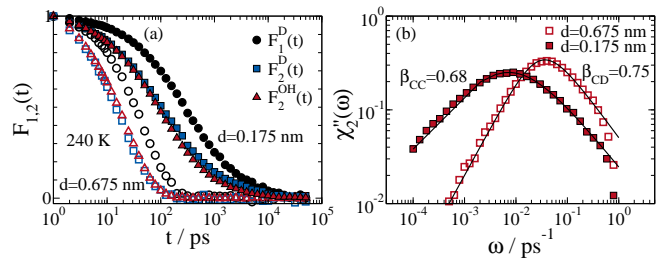}
\caption{Results for water in a neutral pore at 240\,K: (a) Correlation functions $F_1(t)$ and $F_2(t)$ for water in the pore center (open symbols) and at the pore walls (solid symbols). $F_2(t)$ was calculated for both the bond (OH) and dipole (D) vectors. All correlation functions are normalized to the value at $t\!=\!1$\,ps to remove effects from different vibrational contributions.  (b) Imaginary part of the dynamic susceptibilities, $\chi_2^{''}(\omega)$, corresponding to the correlation functions $F_2^{\mathrm{OH}}(t)$. The lines are interpolations with CC and CD functions.}
\label{Fig_4}
\end{figure}

At the end of our computational studies, we calculate experimental observables for water dynamics. First, we compare $F_1(t)$ and $F_2(t)$, which can be obtained from DS and NMR studies, respectively. For isotropic rotational diffusion, i.e., for small-angle jumps, the corresponding correlation times obey the ratio $\tau_1/\tau_2\!=\!3$ \cite{Boehmer01}. In Fig.\ \ref{Fig_4}(a), we see for H2O in CP that $F_1(t)$ indeed decays slower than $F_2(t)$ by about a factor of three not only in the pore center, but also at the pore wall. Hence, effects of large-angle jumps, which occur particularly in the latter region, are too weak to substantially alter this ratio, at least at the studied temperature of 240\,K. Therefore, the ratio of $\tau_1/\tau_2\!=\!3$ needs to be considered when comparing DS and NMR results for the weakly supercooled regime. In addition, inspection of Fig.\ \ref{Fig_4}(a) reveals that correlation functions $F_2(t)$ characterizing the reorientation of the bond and dipole vectors, respectively, agree in the pore center, in harmony with isotropic reorientation in this spatial region, while they differ somewhat at the pore walls. The latter discrepancy is a result of anisotropic reorientation. For example, $\pi$ flips would alter the orientation of the bond vector, but not that of the dipole moment. Since the anisotropy of the reorientation at surfaces becomes more prominent upon cooling, at least in simulations \cite{Vogel09}, it may be relevant to consider this effect when comparing DS and NMR results at low temperatures.

In Fig.\ \ref{Fig_4}(b), we present the imaginary part of the dynamic susceptibility corresponding to $F_2(t)$ for the O--H bond reorientation in various pore region. Evidently, the results for water molecules in the pore center and at the pore wall differ not only with respect to the peak position, reflecting the variation of water mobility across the confinement, but also with regard to the peak shape. Specifically, Cole-Davidson (CD) behavior 
\begin{equation}
\chi_{CD}(\omega)\propto\frac{1}{(1+i\omega\tau_{CD})^{\beta_{CD}}}
\end{equation}
is observed for water in the center, as known from bulk supercooled liquids, while Cole-Cole behavior
\begin{equation}
\chi_{CC}(\omega)\propto\frac{1}{1+(i\omega\tau_{CC})^{\beta_{CC}}}
\end{equation}
is found for water at the wall. Thus, different shapes of the dynamic response functions are another distinguishing feature of water dynamics at rough interfaces and in the bulk liquid.

\subsection{Experimental Studies of Water Dynamics}

In our experimental studies, we first deal with water in protein matrices and later move on to water in silica pores. In both cases, we strive for a characterization of water dynamics in broad temperature ranges and, hence, for suppression of crystallization. For water-protein mixtures, hydration levels $h$ of about 0.3\,g water per 1\,g protein ensure that freezable water is largely absent, while the hydration shells are still filled. For water in silica confinements, pores with diameters of $\sim$2.1\,nm, as found in MCM-41 C10 \cite{Yoshida08}, are suitable to avoid regular freezing while retaining a significant fraction of water molecules, which are not in contact with the silica wall and, thus, possibly exhibit liquid-like behavior. Here, the silica confinements are denoted as C$n$ where $n$ is the number of carbon atoms in the alkyl chain of the precursor molecule, C$_n$H$_{2n+1}$(CH$_3$)$_3$N$^+$Br$^-$. All experimental results show that water dynamics in mixtures and confinements is governed by broad distributions of correlation times $G(\log \tau)$, in harmony with the spatial heterogeneity of water motion in our computational studies. 

\subsubsection{Water Dynamics at Protein Surfaces}\label{Protein}

\begin{figure} 
\includegraphics[width=8.7cm]{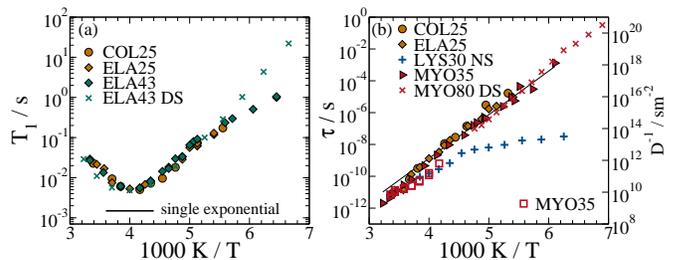}
\caption{(a) $^2$H spin-lattice relaxation times $T_1$ \cite{Vogel08} for heavy water in mixtures with the proteins collagen ($h\!=\!0.25$\,g/g) and elastin ($h\!=\!0.25$\,g/g and $h\!=\!0.43$\,g/g). $T_1$ values calculated based on the spectral density obtained from DS \cite{Lusceac11} are included for comparison. The line marks the minimum value for a single exponential correlation function. (b) Correlation times $\tau_{CC}$ obtained from the $T_1$ values using a CC spectral density. In addition, we show results from analogous analysis for hydrated myoglobin ($h\!=\!0.35$\,g/g) \cite{Lusceac10BBA}, from dielectric spectroscopy on hydrated myoglobin ($h\!=\!0.80$\,g/g) \cite{Swenson06}, and from neutron scattering on hydrated lysozyme ($h\!=\!0.30$\,g/g) \cite{Chen06}. The line marks an Arrhenius law. The open symbols are inverse self-diffusion coefficients $D^{-1}$ of water in a mixture with myoglobin ($h\!=\!0.35$\,g/g) from $^1$H field-gradient NMR \cite{Rosenstihl11}.}
\label{Fig_5}
\end{figure}

Analysis of $^2$H SLR is useful to study rotational motion of heavy water at protein surfaces \cite{Vogel08, Lusceac11, Lusceac10BBA, Vogel10}. Figure \ref{Fig_5}(a) displays temperature-dependent relaxation times $T_1$ for D2O in ELA and COL hydration shells. We see that $T_1$ exhibits a similar minimum for both proteins. From the minimum position, it can be inferred that water reorientation is characterized by a typical correlation time of $\tau_2\!\approx\!1/\omega_0\!\approx\!1\,$ns at 250\,K. From the minimum height, we learn that $F_2(t)$ is not an exponential, consistent with a heterogeneity of the dynamics. Specifically, the measured minimum value is significantly larger than that expected for a Lorentzian shape of the spectral density. The finding that $^2$H SLR is similar for D2O in ELA and COL matrices gives a first hint that water dynamics is comparable at the surfaces of various proteins.

For a determination of correlation times from $T_1$ data, knowledge about the shape of the spectral density $J_2(\omega)$ is required, see Eq.\ (\ref{T1}). DS results, which well agree with NMR data for water-protein systems \cite{Lusceac11, Lusceac10BBA}, provide this information. They revealed that the spectral density for the rotational motion of hydration water has a CC form \cite{Swenson06, Khodadadi08}, consistent with the outcome of our MD simulations. When using the CC spectral density for SLR analysis, the width parameter $\beta_{CC}$ can be determined from the minimum value of $T_1$ and, assuming a temperature-independent width, the time constants $\tau_{CC}$ are available from use of the resulting spectral density in Eq.\ (\ref{T1}). Inspection of Fig.\ \ref{Fig_5}(b) reveals that the correlation times $\tau_{CC}$ are very similar in the hydration shells of COL, ELA, and MYO. The temperature dependence is essentially described by an Arrhenius law with a common activation energy of $E_a\!=\!0.5\,$eV. Thus, the data rule out the existence of a FS crossover, which was reported for the hydration water of lysozyme at 225\,K \cite{Chen06}. Still, the temperature dependence may be slightly higher at ambient temperatures than at cryogenic temperatures, as was reported for myoglobin and lysozyme \cite{Swenson06, Cerveny08, Khodadadi08, Jansson11}.

While $^2$H SLR analysis yields information about rotational motion on local scales, $^1$H FG studies provide insights into translational motion on mesoscopic scales. Self-diffusion coefficients $D$ from the latter approach are included in Fig.\ \ref{Fig_5}(b). The observed values are about an order of magnitude smaller than that for the bulk liquid \cite{Mallamace07}, indicating that substantial water transport occurs in the studied mixtures with hydration levels $h$ of about 0.3\,g/g. The temperature dependence of $D$ is somewhat weaker than that of the correlation times from the SLR analysis, but comparable to that of the time constants from a NS study on hydrated lysozyme \cite{Chen06}. Previously \cite{Rosenstihl11}, we showed that spin relaxation starts to govern the signal decays in $^1$H FG studies on water-preotein mixtures below about 230\,K, marking the limit of the experimental working range. Therefore, kinks of $D(T)$ observed at such temperatures \cite{Mallamace07} do not provide evidence for the existence of a FS transition.         

\begin{figure} 
\includegraphics[width=8.7cm]{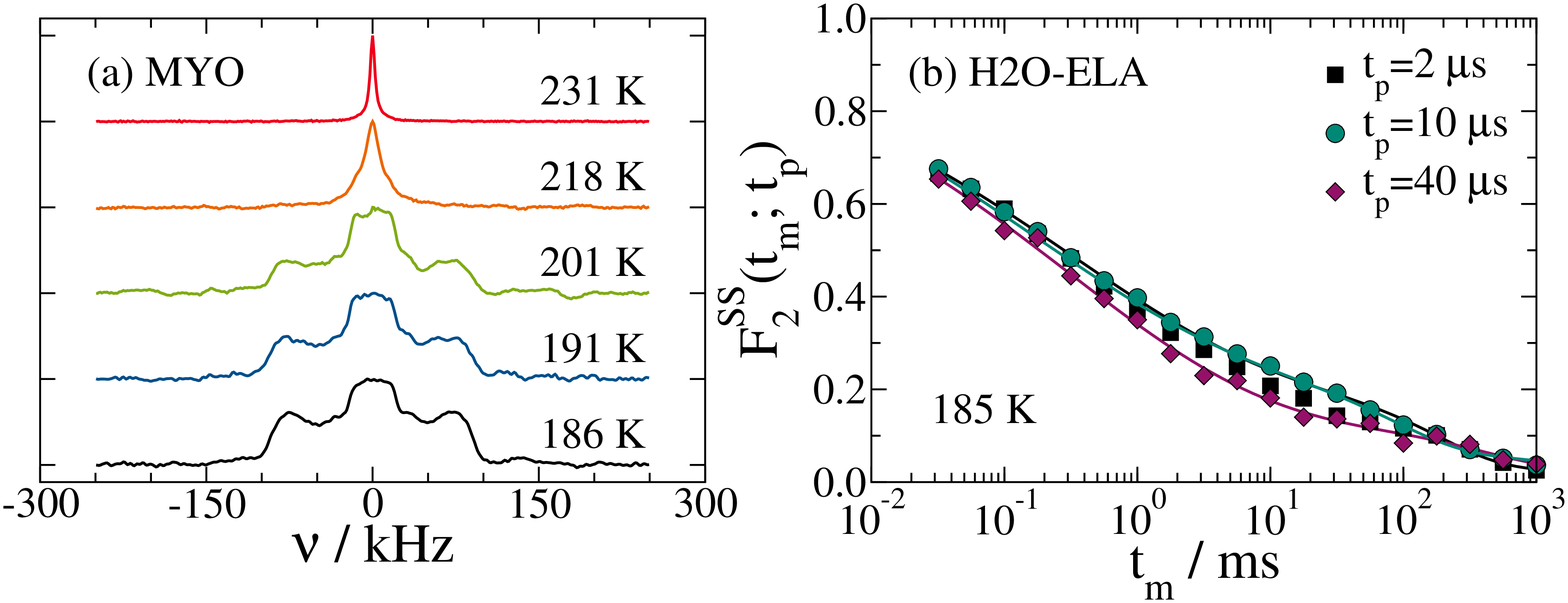}
\caption{$^2$H NMR results for heavy water at proteins surfaces: (a) Spectra for a water-myoglobin mixture ($h\!=\!0.35$\,g/g) at various temperatures. (b) Correlation functions $F_2^{ss}(t_m)$ of a water-elastin mixture ($h\!=\!0.30$\,g/g) at 185\,K. Data for various evolution times are compared. The lines are fits with Eq.\ (\ref{Fit_F2}).}
\label{Fig_6}
\end{figure}

Next, we exploit that $^2$H NMR spectra yield information about the mechanism for water reorientation. For disordered samples, broad and narrow spectra are obtained for slow ($\tau\!\gg\!1/\delta\!\approx\!1\,\mu$s) and fast ($\tau\!\ll\!1/\delta\!\approx\!1\,\mu$s) motions, respectively. The shape of the broad spectrum is given by a Pake pattern as a consequence of the powder average, while the shape of the narrow spectrum is determined by the geometry of the fast motion. Figure \ref{Fig_6}(a) shows $^2$H NMR spectra for a D2O-MYO mixture. Above 230\,K, a narrow Lorentzian line is found, indicating that fast water reorientation together with fast water diffusion lead to an isotropic redistribution of all molecular orientations on the microseconds scale and, hence, average out the orientation dependence of $\omega_Q$. Below 230\,K, the Lorentzian line looses intensity upon cooling while other spectral components appear. Specifically, we observe a Pake spectrum between -160 and +160\,kHz, which continuously grows when the temperature is decreased, and a boxy spectral component, which is most prominent near 200\,K and extends  from ca.\ -20 to +20\,kHz at all temperatures. While the former contribution originates from a growing fraction of static molecules ($\tau\!\gg\!1\,\mu$s) upon cooling, the temperature-independent boxy shape of the latter contribution reveals that a fraction of molecules with $\tau\!\ll\!1\,\mu$s still exists, but their reorientation is no longer isotropic, but anisotropic, resulting in a partial rather than a complete average of the orientation dependence of $\omega_Q$. 

The Lorentzian line finally disappears near 210\,K. At this temperature, the water molecules exhibit a self-diffusion coefficient of $D\!\approx\!10^{-13}$\,m$^2$/s and, hence, an average displacement of 5--10\,$\mathrm{\AA}$ on the microseconds scale of the line-shape experiment, as can be estimated based on an extrapolation of the results in Fig.\ \ref{Fig_5}(b). Thus, the displacements of the water molecules  are smaller than the diameter of the MYO molecules. Therefore, the observation of isotropic and anisotropic water reorientation on the microsecond scale above and below 210\,K, respectively, results because water molecules diffuse through the protein matrix at higher temperatures causing a complete isotropization of molecular orientations, while they stay localized in a certain region of a hydration shell on the experimental time scale at lower temperatures so that an anisotropy of the local reorientation can manifest itself in the line shape.  

In the range 150--200\,K, the $^2$H NMR spectra of the D2O-MYO mixture can be described as a weighted superposition of the 'Box' and Pake patterns. Recently \cite{Lusceac10JPC}, we demonstrated that this observation results from a broad distribution  $G(\log \tau)$ for the anisotropic water reorientation at low temperatures. Specifically, the fast ($\tau\!\ll\!1\,\mu$s) and slow ($\tau\!\gg\!1\,\mu$s) molecules from the distribution give rise to the Box and Pake spectra, respectively, while contributions from molecules with $\tau\!\approx\!1\,\mu$s are negligible. When the temperature is varied, $G(\log \tau)$ shifts and, consequently, the relative intensity of both spectral patterns changes, while their line shapes do not.    

Further insights into the nature of water dynamics at $T\!<\!200\,$K are available from $^2$H STE experiments. In Fig.\ \ref{Fig_6}(b), we show $F_2^{ss}(t_m)$ of a D2O-ELA mixture for various evolution times $t_p$ at 185\,K. Closer analysis \cite{Vogel08} revealed that water motion leads to a decay at short times, while SLR results in additional damping at long times. The loss of correlation due to molecular dynamics is incomplete and nonexponential. Specifically, water reorientation leaves a small, but finite residual correlation \cite{Vogel08} before the onset of relaxation effects, indicating that, though the motion is not isotropic, angular restrictions are not severe. Moreover, water reorientation manifests itself in a stretched exponential decay, $\exp[-(t/\tau)^\beta]$, with a stretching parameter of $\beta\!=\!0.28$ \cite{Vogel08}. Since the angular resolution of the method is higher for longer evolution times, the dependence of the decay time $\tau$ on the value of $t_p$ yields information about jumps angles. While $F_2^{ss}(t_m)$ decays faster for longer evolution times, when the overall reorientation involves successive rotational jumps about small angles, $\tau$ is independent from the value of $t_p$ for rotational jumps about large angles \cite{Fleischer94, Boehmer01}. In Fig.\ \ref{Fig_6}(b), we see that the loss of correlation occurs on a very similar time scale for various values of $t_p$, indicating that the water molecules exhibit large-angle rather than small-angle elementary rotational jumps, i.e., the model of rotational diffusion does not apply, see below. These findings for the motional mechanism, in particular, the observed anisotropy, indicate that, at least at $T\!<\!200\,$K, $^2$H NMR does not probe the $\alpha$ process, but a $\beta$ process of protein hydration water.     

\subsubsection{Water Dynamics in Mesoporous Sililca}\label{Silica}

\begin{figure} 
\includegraphics[width=8.7cm]{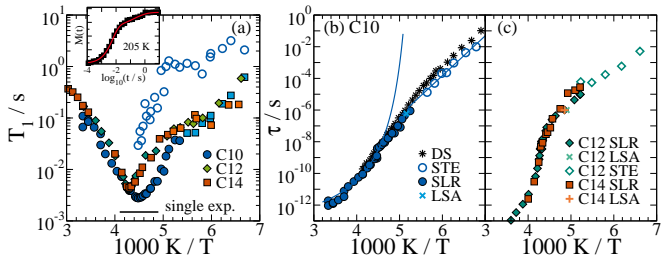}
\caption{$^2$H spin-lattice relaxation times $T_1$ for heavy water in MCM-41 C10, C12, and C14. In all pores, liquid confined water (solid symbols) coexists with solid confined water below ca.\ 225\,K. The $T_1$ values for the latter water fraction are included for the example of C10 (open symbols). The insets show the buildup of the magnetization $M(t)$ for C10 at characteristic temperatures \cite{Sattig14}. The horizontal line marks the minimum value for a single exponential correlation function. Correlation times of water reorientation from spin-lattice relaxation ($\tau_{CC}$), line-shape analysis, and stimulated echo experiments ($\tau_m$) are shown for C10 \cite{Sattig14} in panel (b) and for C12 and C14 in panel (c). For C10, we included fits with Vogel (curved line) and Arrhenius laws (straight line) and data from dielectric spectroscopy \cite{Sjostrom08}.}
\label{Fig_7}
\end{figure}

Next, we perform analogous $^2$H NMR studies on D2O in MCM-41 C10 featuring cylindrical pores with diameters of 2.1\,nm \cite{Yoshida08}. Based on $^2$H SLR results, 3 temperature ranges are distinguishable. While the magnetization $M(t)$ builds up in one step above $\sim$225\,K, it increases in two steps below, see Fig.\ \ref{Fig_7}(a). The latter range is further divided by the finding that single exponential and stretched exponential short-time steps precede the long-time step above and below $\sim$185\,K, respectively. In view of these results, we fit the buildup curves to ($c_l\!+\!c_s\!=\!1$):
\begin{displaymath}
\frac{M(t)}{M(\infty)}=1-c_l\exp\left[-\left(\frac{t}{T_{1,l}}\right)^{\beta_l} \right]-c_s\exp\left[-\left(\frac{t}{T_{1,s}}\right)^{\beta_s} \right]
\end{displaymath}
For reasons to be explained, we utilize the indices $l$ (liquid) and $s$ (solid) to discriminate between the relaxation steps. The relaxation times $T_{1,l}$ and $T_{1,s}$ are displayed in Fig.\ \ref{Fig_7}(a). For the short-time step, light blue and dark blue symbols are used to show that $\beta_l\!=\!1$ above $\sim$185\,K, while $\beta_l\!<\!1$ below. For the long-time step, $\beta_s\!\approx\!0.6$ reveals nonexponential relaxation at all temperatures.

The shape of the buildup curves provides valuable information about the confined water. For the following arguments, it is important to recall that spatial heterogeneity is a key feature of water dynamics in mixtures and confinements. In general, in $^2$H NMR, the corresponding distribution of correlation times $\tau$ results in a distribution of relaxation times $T_1$ and, hence, in a nonexponentiality of $M(t)$. This argument is valid when the correlation times of the molecules are unchanged during the buildup of the magnetization, which usually occurs on a much longer time scale than the molecular reorientation, while it does not apply when an exchange of $\tau$ values averages over a distribution of $T_1$ values and, thus, reconstitutes exponential relaxation. The former scenario is expected when molecular diffusion is quenched in solids, whereas the latter scenario applies to liquids where the molecules explore different local environments in the course of time. 

Thus, above 225\,K, the observation of monoexponential $^2$H SLR indicates that the water molecules exchange their correlation times on the milliseconds time scale of the buildup of $M(t)$ and, hence, sample a substantial part of the pore volume, as expected for a liquid. Below 225\,K, the existence of two relaxation steps provides clear evidence that there are dynamically distinguishable water fractions that do not exchange molecules until the buildup of magnetization is complete. In this temperature range, the nonexponentiality of the long-time step shows that the associated water fraction does not explore different local environments on the time scale of $T_1$ and, thus, these molecules form a solid. The crossover between exponential and nonexponential behavior of the short-time step near 185\,K implies that the corresponding water fraction continues to explore a relevant part of the pore volume and, hence, stays liquid above this temperature, while molecular diffusion becomes too slow to restore ergodicity on the time scale of the buildup process below, resembling the situation for supercooled liquids at $T_g$ \cite{Schnauss90, Boehmer01}. Consistently, $T_1(T)$ exhibits a kink at $\sim$185\,K, as usually observed for bulk and confined liquids undergoing a glass transition \cite{Lusceac04}. Altogether, the observations indicate that the confined water splits into liquid and solid fractions upon cooling through 225\,K, explaining our nomenclature for the relaxation steps. Moreover, the findings imply that the liquid fraction undergoes a confinement-affected glass transition near 185\,K.

The relaxation times $T_{1,l}$ provide access to the correlation times of molecular reorientation in liquid water. This analysis can performed in analogy with the above $^2$H SLR studies of water dynamics at protein surfaces. In particular, it is possible to exploit information from DS work \cite{Sjostrom08} and use a CC spectral density. In Fig.\ \ref{Fig_5}(b), we observe that the resulting correlation times $\tau_{CC}$ exhibit a mild change in the temperature dependence near 225\,K \cite{Sattig14}. The correlation times $\tau_{CC}$ obtained from analogous analysis for D2O in MCM-41 C12 and C14 are presented in Fig.\ \ref{Fig_5}(c). We find that the dynamical crossover is more prominent for water in these somewhat larger pores with diameters up 2.9\,nm. The different sharpness stems from the findings that, below 225\,K, the temperature dependence is comparable in all confinements, while it is stronger and, hence, more bulk-like, in the wider C12 and C14 pores than in the narrower C10 pores above this temperature, consistent with the pore-size dependence of water dynamics in our simulation studies. Interestingly, the observed change of water dynamics at 225\,K is accompanied by the emergence of solid water in all confinements. Hence, it does not necessarily indicate a FS crossover related to a LL transition, see below.          

\begin{figure} 
\includegraphics[width=8.7cm]{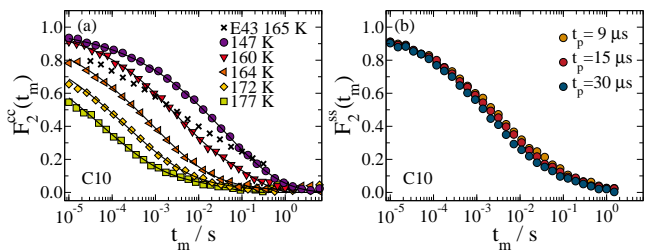}
\caption{$^2$H stimulated-echo decays for heavy water in MCM-41 C10. (a) $F_2^{cc}(t_m)$ at various temperatures for $t_p\!=\!9\,\mu$s \cite{Sattig14} together with results for hydrated elastin ($h\!=\!0.43$\,g/g) at 165\,K \cite{Vogel08}. The lines are fits with Eq.\ (\ref{Fit_F2}). (b) $F_2^{ss}(t_m)$ for various evolution times at 160\,K.}
\label{Fig_8}
\end{figure}

Different scenarios \cite{Stanley10, Mallamace12, Cerveny04, Swenson06, Hedstrom07, Cerveny08, Khodadadi08, Pawlus08, Bruni11, Gainaru09, Lusceac11, Vogel08, Sattig14, Doster10, Ngai10, Capaccioli11, Zanotti05, Liu05, Chen06, Yoshida08, Kumar06, Gallo10, Doster10, Limmer11} were proposed to rationalize changes in the temperature dependence of water dynamics. To discriminate between various conjectures, we ascertain the nature of water motion below the crossover temperature, e.g., to determine whether the $\alpha$ or $\beta$ process of water is observed in this range. Recently \cite{Sattig14}, we investigated $^2$H NMR spectra to obtain insights into the mechanism for water reorientation. We found that a Lorentzian contribution exists down to $\sim$185\,K, indicating that at least a fraction of water molecules shows sufficiently fast isotropic reorientation, consistent with an exploration of a substantial part of the pore volume, as was above inferred from $\beta_l\!=\!1$. In the following, we use $^2$H STE experiments to investigate water reorientation below 185\,K. In particular, we perform partially relaxed STE measurements, which exploit the different SLR of the water species, $T_{1,l}\!\ll\!T_{1,s}$, to single out contributions of the water faction that stays liquid at 225\,K and to suppress those of the water fraction that becomes solid at this temperature \cite{Sattig14}.

Figure \ref{Fig_8}(a) shows $F_2^{cc}(t_m)$ for D2O in MCM-41 C10 at various temperatures, as obtained from partially relaxed measurements. It is evident that strongly nonexponential decays shift to lower temperatures upon cooling. For a quantitative analysis, we fit the normalized data to
\begin{equation}
\left[(1\!-\! F_\infty ) \exp \left[ -\left(\frac{t_m}{\tau_K}\right)^{\beta_K}  \right] \!+\! F_\infty \right]  \Phi_l (t_m) 
\label{Fit_F2}
\end{equation}         
Hence, we use a stretched exponential to describe the signal decay due to water reorientation and utilize a residual correlation $F_\infty$ to consider possible anisotropy. Furthermore, we exploit that, in partially relaxed experiments, SLR damping is described by $\Phi_l(t_m)$, which is obtained from SLR analysis. From the fit results, we calculate mean logarithmic correlation times $\tau_m$ according to \cite{Zorn02} 
\begin{equation}\label{mltau}
\langle \ln \tau \rangle \equiv \ln \tau_m = \ln \tau_K + (1-\frac{1}{\beta_K})\mathrm{Eu}
\end{equation}
Here, $\mathrm{Eu}\!\approx\!0.58$ is Euler's constant. In Fig.\ \ref{Fig_7}(b), $\tau_m$ from STE analysis is shown together with $\tau_{CC}$ from SLR analysis, which is also a mean logarithmic correlation time due to the symmetric shape of the CC distribution. We see that the temperature dependence obtained from the STE experiments below 185\,K is somewhat weaker than that resulting from the SLR approach above, in nice agreement with findings in DS works \cite{Sjostrom08,Bruni11}. Thus, the temperature dependence changes not only at $\sim$225\,K, but also at $\sim$185\,K, where the liquid water fraction may undergo a glass transition, as aforementioned. Inspection of Fig. \ref{Fig_7}(c) reveals that the same conclusions can be drawn when comparing SLR and STE data for D2O in MCM-41 C12. Below 185\,K, $\tau_m$ follows an Arrhenius law with an activation energy of $E_a\!\approx\!0.5\,$eV, which is characteristic for low-temperature water reorientation at various types of surfaces \cite{Cerveny04, Cerveny08}. Yet, direct comparison reveals that the rotational correlation functions decay somewhat faster and less stretched for water in silica pores than for water at elastin surfaces, see Fig.\ \ref{Fig_8}(a). 

\begin{figure} 
\includegraphics[width=8.7cm]{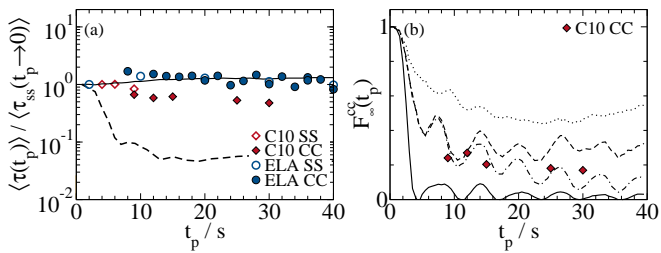}
\caption{Evolution-time dependence of $F_2^{cc}(t_m)$ and $F_2^{ss}(t_m)$ for heavy water in silica pores (MCM-41 C10 \cite{Sattig14}) and at elastin surfaces ($h\!=\!0.43$\,g/g \cite{Vogel08}): (a) correlation times and (b) residual correlations. In panel (a), the experimental data, which are normalized by the time constant resulting from $F_2^{ss}(t_m,t_p\!\rightarrow\!0)\!\approx\!F_2(t_m)$, are compared with simulation results for tetrahedral jumps (solid line) and isotropic $10^\circ$ jumps (dashed line). In panel (b), the measured data are contrasted with expectations for isotropic reorientation (solid line), $180^\circ$ jumps about the molecular symmetry axis (dotted line), and tetrahedral jumps. In the latter case, we distinguish between exact tetrahedral jumps (dashed line) and distorted ($\pm3^\circ$) tetrahedral jumps (dash-dotted line).}
\label{Fig_9}
\end{figure}

In analogy with our strategy for protein matrices, we exploit for silica confinements that the mechanism for low-temperature water reorientation can be determined when analyzing the dependence of $F_2^{cc}(t_m)$ and $F_2^{ss}(t_m)$ on the evolution time $t_p$. For D2O in MCM-41 C10 at 160\,K, a weak evolution-time dependence of $F_2^{ss}(t_m)$ is evident from Fig.\ \ref{Fig_8}(b). This observation is confirmed when fitting the decays to Eq.\ (\ref{Fit_F2}). In Fig.\ \ref{Fig_9}(a), we see that the mean time constants hardly depend on the value of $t_p$, consistent with results from $F_2^{cc}(t_m)$. These findings reveal that water reorientation in silica matrices at low temperatures involves elementary jumps about large angles of the order of the tetrahedral angle, in harmony with our results for water at protein surfaces. By contrast, a strong evolution-time dependence of the time constants is characteristic for the $\alpha$ process of supercooled liquids, indicating a high relevance of jumps about small angles \cite{Boehmer01, Fleischer94, Boehmer98, Hinze98, Wachner99}. In Fig.\ \ref{Fig_9}(b), we observe a finite residual correlation $F_\infty^{cc}\!\approx\!0.2$. Hence, water reorientation is anisotropic and does not destroy all correlation. Further insights are available from comparison with expectations for various motional models. The observed residual correlation is significantly higher and lower than the expectations for isotropic reorientation and $\pi$ flips, respectively, while it is in rough agreement with a tetrahedral jump, in particular, when we allow for mild distortions ($\pm3^\circ$). However, a unique determination of the motional geometry is not possible due to imperfections in the suppression of the solid water species  in partially relaxed measurements and to an interference of SLR damping, being even more problematic for $F_\infty^{ss}$, which is, therefore, not discussed. 

Altogether, the observed changes in the temperature dependence and the motional mechanism reveal that the $\beta$ process rather than the $\alpha$ process governs the experimental findings for water in mesoporous silica well below 225\,K.       

\subsection{Interplay of Water and Protein Dynamics}

\begin{figure} 
\includegraphics[width=8.7cm]{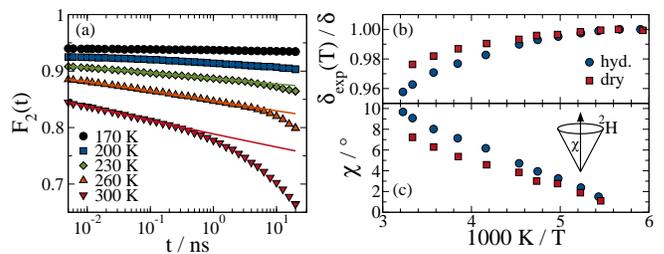}
\caption{(a) Correlation functions $F_2(t)$ for the reorientation of carbonyl groups in hydrated elastin ($h\!=\!0.3$\,g/g) from MD simulations \cite{Kampf12}. The lines indicate power laws. (b) Normalized anisotropy parameter $\delta_{\mathrm{exp}}/\delta$ characterizing $^2$H NMR Pake spectra of dry and hydrated ($h\!=\!0.3\,$g/g) C-phycocyanin at various temperatures \cite{Kampf14}. (c) Semiopening angles $\chi$ obtained within the shown cone model from the anisotropy parameters using Eq.\ (\ref{delta}) \cite{Kampf14}.}
\label{Fig_10}
\end{figure}

Finally, we use MD simulations and NMR experiments to ascertain protein backbone dynamics for a moderate hydration level of $h\!=\!0.3$\,g/g. For such an amount of water, relevant internal motion is already activated, while overall tumbling motion is still suppressed. 

In Fig.\ \ref{Fig_10}(a), we display correlation functions $F_2(t)$ from simulations of a H2O-ELA mixture at various temperatures, which describe the reorientation of the carbonyl groups in the peptide bonds \cite{Kampf12}. It can be seen that the rotational motion of the protein backbone manifests itself in an anomalous loss, i.e., in power-law (PL) or logarithmic-like (LG) decays, which extend from the picoseconds to the nanoseconds regimes until the amplitude of the decays vanishes upon cooling at 170--190\,K \cite{Kampf12}. Similar findings were reported for other proteins \cite{Lagi09, Kampf12}, implying that the existence of PL/LG decays does not rely on specific secondary structures. Various approaches were employed to rationalize the anomalous protein dynamics \cite{Frauenfelder09, Kampf12, Lagi09, Kneller04}. For the H2O-ELA mixture, the simulation results can be described in the framework of a fractional Ornstein-Uhlenbeck process \cite{Kampf12}, i.e., when using a fractional Fokker-Planck approach \cite{Metzler00} to model anomalous diffusion in a harmonic potential caused by neighboring particles \cite{Kampf12}. Then, the temperature dependence of the anomalous decays solely results from decreasing amplitudes of position fluctuations within the local cages upon cooling.

In Figs.\ \ref{Fig_10}(b) and (c), we present results from $^2$H NMR approaches to the backbone dynamics of C-phycocyanin (CPC). In the studied range 150--300\,K, we find that a variation of temperature leads to minor changes of the line width, but not to major modifications of the line shape, i.e., there is no collapse of the Pake pattern. The weak line narrowing can be characterized when extracting the anisotropy parameter $\delta_\mathrm{exp}$ from the experimental spectra at various temperatures. In panel (b), it is evident that, for both dry and hydrated CPC, the static limit $\delta_\mathrm{exp}\!=\!\delta$ is observed below 175--185\,K, indicating an absence of backbone dynamics with $\tau\!<\!1\,\mu$s. Above this range, an increase of temperature results in a continuous decrease of the line width. This motional narrowing is weak but more prominent for hydrated CPC than for dry CPC. Consequently, the backbone motion is highly restricted and water coupled.

Considering also results from $^2$H SLR and STE studies on CPC, we showed in recent work \cite{Kampf14} that the temperature dependence of the line width is due to a change of the geometry rather than to a variation of the rate of backbone motion, consistent with the above MD data. Specifically, the analysis revealed that all backbone deuterons exhibit rotational motions on time scales faster than microseconds, which become more restricted upon cooling. In such case, angular amplitudes can be determined from the observed anisotropy parameters $\delta_\mathrm{exp}$. Assuming jumps on the surface of a cone, the semi-opening angle of the cone $\chi$ is obtained from comparison of the observed and static line widths according to
\begin{equation}\label{delta}
\delta_\mathrm{exp}=\frac{\delta}{2}(3\cos^2\chi-1)
\end{equation}
Fig.\ \ref{Fig_10}(c) compares results for dry and hydrated CPC. At room temperature, addition of hydration water results in an increase of the semi-opening angle from $\chi\!=\!7^\circ$ to $\chi\!=\!10^\circ$. Upon cooling, the amplitude of the reorientation continuously decreases until it vanishes near 175-185\,K for both the dry and hydrated proteins.  

Thus, the following picture emerges when combining the results of our computational and experimental studies: The protein backbone shows fast dynamics, which manifests itself in PL or LG decays of correlation functions and, hence, has no characteristic time scale, resembling the nearly-constant loss phenomenon of disordered systems. Upon cooling the amplitude of the backbone motion continuously decreases until it disappears at 170--190\,K, reflecting a dynamical transition of the protein. Interestingly, this dynamical transition of proteins and the glass transition of water at surfaces occur in the same temperature range, possibly providing further evidence for the importance of water-protein couplings.      

\section{Conclusion}

Our combined computational and experimental results revealed substantial differences between water dynamics at an interface and in the bulk. Specifically, interfacial water is significantly slower and less fragile than bulk water. Moreover, jump processes are more important for structural relaxation in the former than in the latter case and Cole-Cole behavior develops from Cole-Davidson behavior when approaching an interface. These changes do not require attractive interactions, but sufficient roughness and rigidity of the surface so that the surface atoms provide a static contribution to the energy landscape for neighboring water molecules. Water dynamics at such interfaces resembles other dynamical processes in solid matrices, e.g., plasticizer dynamics in polymers \cite{Boehmer01, Vogel05} or ion dynamics in glasses \cite{Boehmer07, Brinkmann10}. The interfacial region covers 2--4 water layers, where the range of the surface effect mildly increases upon cooling and weakly depends on the surface chemistry.  

In weakly hydrated proteins and in partially filled or very narrow confinements, no bulk water exists, but all water molecules reside at an interface. At rigid surfaces, water dynamics obeys an Arrhenius law with an activation energy of $E_a\!=\!0.5$\,eV \cite{Cerveny04, Cerveny08, Gainaru09}. Mild deviations from this Arrhenius law can occur when the matrix flexibility increases, e.g., upon heating through a glass transition \cite{Cerveny08, Ngai10, Capaccioli11}. While the temperature dependence is universal, the absolute value of the correlation time is smaller for higher water fractions \cite{Gainaru09, Sjostrom10}. Our studies revealed that, at essentially rigid interfaces and at sufficiently low temperatures, the elementary steps of water dynamics are large-angle jumps with a mild anisotropy, reflecting interactions with the respective surface. In these situations, the imposed energy landscape hinders cooperative water rearrangements so that the $\beta$ process rather than the $\alpha$ process of water is observed. 

The dynamical scenario is more complex for moderately hydrated proteins or completely filled pores with diameters of a few nanometers. In such systems, the water mobility strongly differs in various confinement regions. At sufficiently high temperatures, the water molecules explore the whole confinement in the course of time and, hence, switch between bulk-like and interface-dominated dynamical states. Then, the temperature dependence of water dynamics does no longer obey an Arrhenius law. Thereby, the fragility depends on the confinement size. When a fraction of water crystallizes or vitrifies \cite{Swenson10} in the center of the confinement upon cooling, e.g., at 220--230\,K in our case of silica pores with diameters of 2.1--2.9\,nm, the motion of the remaining fraction of liquid water becomes restricted to narrow regions near the matrix, putting severe limits to cooperativity. Consequently, bulk-like structural relaxation is replaced by interface-dominated water motion. Moreover, single-particle correlation functions cease to represent the $\alpha$ process and start to probe the $\beta$ process. Therefore, kinks in temperature-dependent correlation times of confined water at 220--230\,K do not necessarily yield evidence for the existence of a FS crossover related to a LL transition of the bulk liquid. At sufficiently low temperatures, the $\beta$ process, as a consequence of its large amplitude, dominates the decays of single-particle correlation functions, rendering observation of a possible $\alpha$ process difficult in standard experiments. Nevertheless, we found that structural rearrangements of interfacial water continue down to a glass-transition like event at $\sim$185\,K. However, due to strong effects of interfaces on the underlying water dynamics, this value is not to be identified with the glass transition temperature of bulk water. 

For hydrated proteins, we observed highly restricted backbone motion, which manifests itself in power-law or logarithmic-like correlation functions and, hence, has no characteristic time. The amplitude of this motion decreases upon cooling and vanishes at 170--190\,K. Hence, this dynamical transition of proteins occurs in same temperature range as a glass transition of water at interfaces, suggesting a dynamical coupling of these protein and water processes.  

\section{Acknowledgment}
We thank Jan Swenson (Chalmers University) and Gerd Buntkowsky (Technische Universit\"at Darmstadt) for providing us with mesoporous silica and the Deutsche Forschungsgemeinschaft for funding through Grants No.\ Vo-905/8-1 and Vo-905/9-1.    

\section*{References}

\bibliography{Biblio_Vogel}

\begin{thebibliography}{72}%
\makeatletter
\providecommand \@ifxundefined [1]{%
 \@ifx{#1\undefined}
}%
\providecommand \@ifnum [1]{%
 \ifnum #1\expandafter \@firstoftwo
 \else \expandafter \@secondoftwo
 \fi
}%
\providecommand \@ifx [1]{%
 \ifx #1\expandafter \@firstoftwo
 \else \expandafter \@secondoftwo
 \fi
}%
\providecommand \natexlab [1]{#1}%
\providecommand \enquote  [1]{``#1''}%
\providecommand \bibnamefont  [1]{#1}%
\providecommand \bibfnamefont [1]{#1}%
\providecommand \citenamefont [1]{#1}%
\providecommand \href@noop [0]{\@secondoftwo}%
\providecommand \href [0]{\begingroup \@sanitize@url \@href}%
\providecommand \@href[1]{\@@startlink{#1}\@@href}%
\providecommand \@@href[1]{\endgroup#1\@@endlink}%
\providecommand \@sanitize@url [0]{\catcode `\\12\catcode `\$12\catcode
  `\&12\catcode `\#12\catcode `\^12\catcode `\_12\catcode `\%12\relax}%
\providecommand \@@startlink[1]{}%
\providecommand \@@endlink[0]{}%
\providecommand \url  [0]{\begingroup\@sanitize@url \@url }%
\providecommand \@url [1]{\endgroup\@href {#1}{\urlprefix }}%
\providecommand \urlprefix  [0]{URL }%
\providecommand \Eprint [0]{\href }%
\providecommand \doibase [0]{http://dx.doi.org/}%
\providecommand \selectlanguage [0]{\@gobble}%
\providecommand \bibinfo  [0]{\@secondoftwo}%
\providecommand \bibfield  [0]{\@secondoftwo}%
\providecommand \translation [1]{[#1]}%
\providecommand \BibitemOpen [0]{}%
\providecommand \bibitemStop [0]{}%
\providecommand \bibitemNoStop [0]{.\EOS\space}%
\providecommand \EOS [0]{\spacefactor3000\relax}%
\providecommand \BibitemShut  [1]{\csname bibitem#1\endcsname}%
\let\auto@bib@innerbib\@empty
%</preamble>
\bibitem [{\citenamefont {Debenedetti}(2003)}]{Debenedetti03}%
  \BibitemOpen
  \bibfield  {author} {\bibinfo {author} {\bibfnamefont {P.~G.}\ \bibnamefont
  {Debenedetti}},\ }\href@noop {} {\bibfield  {journal} {\bibinfo  {journal}
  {J. Phys.: Condens. Matter}\ }\textbf {\bibinfo {volume} {15}},\ \bibinfo
  {pages} {R1669} (\bibinfo {year} {2003})}\BibitemShut {NoStop}%
\bibitem [{\citenamefont {Bagchi}(2005)}]{Bagchi05}%
  \BibitemOpen
  \bibfield  {author} {\bibinfo {author} {\bibfnamefont {B.}~\bibnamefont
  {Bagchi}},\ }\href@noop {} {\bibfield  {journal} {\bibinfo  {journal} {Chem.
  Rev.}\ }\textbf {\bibinfo {volume} {105}},\ \bibinfo {pages} {3197} (\bibinfo
  {year} {2005})}\BibitemShut {NoStop}%
\bibitem [{\citenamefont {Frauenfelder}\ \emph {et~al.}(2009)\citenamefont
  {Frauenfelder}, \citenamefont {Chen}, \citenamefont {Berendzen},
  \citenamefont {Fenimore}, \citenamefont {Jansson}, \citenamefont {McMahon},
  \citenamefont {Stroe}, \citenamefont {Swenson},\ and\ \citenamefont
  {Young}}]{Frauenfelder09}%
  \BibitemOpen
  \bibfield  {author} {\bibinfo {author} {\bibfnamefont {H.}~\bibnamefont
  {Frauenfelder}}, \bibinfo {author} {\bibfnamefont {G.}~\bibnamefont {Chen}},
  \bibinfo {author} {\bibfnamefont {J.}~\bibnamefont {Berendzen}}, \bibinfo
  {author} {\bibfnamefont {P.~W.}\ \bibnamefont {Fenimore}}, \bibinfo {author}
  {\bibfnamefont {H.}~\bibnamefont {Jansson}}, \bibinfo {author} {\bibfnamefont
  {B.~H.}\ \bibnamefont {McMahon}}, \bibinfo {author} {\bibfnamefont {I.~R.}\
  \bibnamefont {Stroe}}, \bibinfo {author} {\bibfnamefont {J.}~\bibnamefont
  {Swenson}}, \ and\ \bibinfo {author} {\bibfnamefont {R.~D.}\ \bibnamefont
  {Young}},\ }\href@noop {} {\bibfield  {journal} {\bibinfo  {journal} {Proc.
  Natl. Acad. Sci. USA}\ }\textbf {\bibinfo {volume} {106}},\ \bibinfo {pages}
  {5129} (\bibinfo {year} {2009})}\BibitemShut {NoStop}%
\bibitem [{\citenamefont {Angell}(2002)}]{Angell02}%
  \BibitemOpen
  \bibfield  {author} {\bibinfo {author} {\bibfnamefont {C.~A.}\ \bibnamefont
  {Angell}},\ }\href@noop {} {\bibfield  {journal} {\bibinfo  {journal} {Chem.
  Rev.}\ }\textbf {\bibinfo {volume} {102}},\ \bibinfo {pages} {2627} (\bibinfo
  {year} {2002})}\BibitemShut {NoStop}%
\bibitem [{\citenamefont {Angell}(2008)}]{Angell08}%
  \BibitemOpen
  \bibfield  {author} {\bibinfo {author} {\bibfnamefont {C.~A.}\ \bibnamefont
  {Angell}},\ }\href@noop {} {\bibfield  {journal} {\bibinfo  {journal}
  {Science}\ }\textbf {\bibinfo {volume} {319}},\ \bibinfo {pages} {582}
  (\bibinfo {year} {2008})}\BibitemShut {NoStop}%
\bibitem [{\citenamefont {Ngai}\ \emph {et~al.}(2013)\citenamefont {Ngai},
  \citenamefont {Capaccioli},\ and\ \citenamefont {Paciaroni}}]{Ngai13}%
  \BibitemOpen
  \bibfield  {author} {\bibinfo {author} {\bibfnamefont {K.~L.}\ \bibnamefont
  {Ngai}}, \bibinfo {author} {\bibfnamefont {S.}~\bibnamefont {Capaccioli}}, \
  and\ \bibinfo {author} {\bibfnamefont {A.}~\bibnamefont {Paciaroni}},\
  }\href@noop {} {\bibfield  {journal} {\bibinfo  {journal} {Chem. Phys.}\
  }\textbf {\bibinfo {volume} {424}},\ \bibinfo {pages} {37} (\bibinfo {year}
  {2013})}\BibitemShut {NoStop}%
\bibitem [{\citenamefont {Poole}\ \emph {et~al.}(1992)\citenamefont {Poole},
  \citenamefont {Sciortino}, \citenamefont {Essmann},\ and\ \citenamefont
  {Stanley}}]{Poole92}%
  \BibitemOpen
  \bibfield  {author} {\bibinfo {author} {\bibfnamefont {P.~H.}\ \bibnamefont
  {Poole}}, \bibinfo {author} {\bibfnamefont {F.}~\bibnamefont {Sciortino}},
  \bibinfo {author} {\bibfnamefont {U.}~\bibnamefont {Essmann}}, \ and\
  \bibinfo {author} {\bibfnamefont {H.~E.}\ \bibnamefont {Stanley}},\
  }\href@noop {} {\bibfield  {journal} {\bibinfo  {journal} {Nature}\ }\textbf
  {\bibinfo {volume} {360}},\ \bibinfo {pages} {324} (\bibinfo {year}
  {1992})}\BibitemShut {NoStop}%
\bibitem [{\citenamefont {Mishima}\ and\ \citenamefont
  {Stanley}(1998)}]{Mishima98}%
  \BibitemOpen
  \bibfield  {author} {\bibinfo {author} {\bibfnamefont {O.}~\bibnamefont
  {Mishima}}\ and\ \bibinfo {author} {\bibfnamefont {H.~E.}\ \bibnamefont
  {Stanley}},\ }\href@noop {} {\bibfield  {journal} {\bibinfo  {journal}
  {Science}\ }\textbf {\bibinfo {volume} {396}},\ \bibinfo {pages} {329}
  (\bibinfo {year} {1998})}\BibitemShut {NoStop}%
\bibitem [{\citenamefont {Ito}\ \emph {et~al.}(1999)\citenamefont {Ito},
  \citenamefont {Moynihan},\ and\ \citenamefont {Angell}}]{Ito99}%
  \BibitemOpen
  \bibfield  {author} {\bibinfo {author} {\bibfnamefont {K.}~\bibnamefont
  {Ito}}, \bibinfo {author} {\bibfnamefont {C.~T.}\ \bibnamefont {Moynihan}}, \
  and\ \bibinfo {author} {\bibfnamefont {C.~A.}\ \bibnamefont {Angell}},\
  }\href@noop {} {\bibfield  {journal} {\bibinfo  {journal} {Nature}\ }\textbf
  {\bibinfo {volume} {398}},\ \bibinfo {pages} {492} (\bibinfo {year}
  {1999})}\BibitemShut {NoStop}%
\bibitem [{\citenamefont {Stanley}\ \emph {et~al.}(2010)\citenamefont
  {Stanley}, \citenamefont {Buldyrev}, \citenamefont {Franzese}, \citenamefont
  {Kumar}, \citenamefont {Mallamace}, \citenamefont {Mazza}, \citenamefont
  {Stokely},\ and\ \citenamefont {Xu}}]{Stanley10}%
  \BibitemOpen
  \bibfield  {author} {\bibinfo {author} {\bibfnamefont {H.~E.}\ \bibnamefont
  {Stanley}}, \bibinfo {author} {\bibfnamefont {S.~V.}\ \bibnamefont
  {Buldyrev}}, \bibinfo {author} {\bibfnamefont {G.}~\bibnamefont {Franzese}},
  \bibinfo {author} {\bibfnamefont {P.}~\bibnamefont {Kumar}}, \bibinfo
  {author} {\bibfnamefont {F.}~\bibnamefont {Mallamace}}, \bibinfo {author}
  {\bibfnamefont {M.~G.}\ \bibnamefont {Mazza}}, \bibinfo {author}
  {\bibfnamefont {K.}~\bibnamefont {Stokely}}, \ and\ \bibinfo {author}
  {\bibfnamefont {L.}~\bibnamefont {Xu}},\ }\href@noop {} {\bibfield  {journal}
  {\bibinfo  {journal} {J. Phys.: Condens. Matter}\ }\textbf {\bibinfo {volume}
  {22}},\ \bibinfo {pages} {284101} (\bibinfo {year} {2010})}\BibitemShut
  {NoStop}%
\bibitem [{\citenamefont {Mallamace}\ \emph {et~al.}(2012)\citenamefont
  {Mallamace}, \citenamefont {Corsaro}, \citenamefont {Baglioni}, \citenamefont
  {Fratini},\ and\ \citenamefont {Chen}}]{Mallamace12}%
  \BibitemOpen
  \bibfield  {author} {\bibinfo {author} {\bibfnamefont {F.}~\bibnamefont
  {Mallamace}}, \bibinfo {author} {\bibfnamefont {C.}~\bibnamefont {Corsaro}},
  \bibinfo {author} {\bibfnamefont {P.}~\bibnamefont {Baglioni}}, \bibinfo
  {author} {\bibfnamefont {E.}~\bibnamefont {Fratini}}, \ and\ \bibinfo
  {author} {\bibfnamefont {S.-H.}\ \bibnamefont {Chen}},\ }\href@noop {}
  {\bibfield  {journal} {\bibinfo  {journal} {J. Phys.: Condens. Matter}\
  }\textbf {\bibinfo {volume} {24}},\ \bibinfo {pages} {064103} (\bibinfo
  {year} {2012})}\BibitemShut {NoStop}%
\bibitem [{\citenamefont {Zanotti}\ \emph {et~al.}(2005)\citenamefont
  {Zanotti}, \citenamefont {Bellissent-Funel},\ and\ \citenamefont
  {Chen}}]{Zanotti05}%
  \BibitemOpen
  \bibfield  {author} {\bibinfo {author} {\bibfnamefont {J.-M.}\ \bibnamefont
  {Zanotti}}, \bibinfo {author} {\bibfnamefont {M.-C.}\ \bibnamefont
  {Bellissent-Funel}}, \ and\ \bibinfo {author} {\bibfnamefont {S.-H.}\
  \bibnamefont {Chen}},\ }\href@noop {} {\bibfield  {journal} {\bibinfo
  {journal} {Europhys. Lett.}\ }\textbf {\bibinfo {volume} {71}},\ \bibinfo
  {pages} {91} (\bibinfo {year} {2005})}\BibitemShut {NoStop}%
\bibitem [{\citenamefont {Liu}\ \emph {et~al.}(2005)\citenamefont {Liu},
  \citenamefont {Chen}, \citenamefont {Faraone}, \citenamefont {Yen},\ and\
  \citenamefont {Mou}}]{Liu05}%
  \BibitemOpen
  \bibfield  {author} {\bibinfo {author} {\bibfnamefont {L.}~\bibnamefont
  {Liu}}, \bibinfo {author} {\bibfnamefont {S.-H.}\ \bibnamefont {Chen}},
  \bibinfo {author} {\bibfnamefont {A.}~\bibnamefont {Faraone}}, \bibinfo
  {author} {\bibfnamefont {C.-W.}\ \bibnamefont {Yen}}, \ and\ \bibinfo
  {author} {\bibfnamefont {C.-Y.}\ \bibnamefont {Mou}},\ }\href@noop {}
  {\bibfield  {journal} {\bibinfo  {journal} {Phys. Rev. Lett.}\ }\textbf
  {\bibinfo {volume} {95}},\ \bibinfo {pages} {117802} (\bibinfo {year}
  {2005})}\BibitemShut {NoStop}%
\bibitem [{\citenamefont {Chen}\ \emph {et~al.}(2006)\citenamefont {Chen},
  \citenamefont {Liu}, \citenamefont {Fratini}, , \citenamefont {Baglioni},
  \citenamefont {Faraone},\ and\ \citenamefont {Mamontov}}]{Chen06}%
  \BibitemOpen
  \bibfield  {author} {\bibinfo {author} {\bibfnamefont {S.-H.}\ \bibnamefont
  {Chen}}, \bibinfo {author} {\bibfnamefont {L.}~\bibnamefont {Liu}}, \bibinfo
  {author} {\bibfnamefont {E.}~\bibnamefont {Fratini}}, , \bibinfo {author}
  {\bibfnamefont {P.}~\bibnamefont {Baglioni}}, \bibinfo {author}
  {\bibfnamefont {A.}~\bibnamefont {Faraone}}, \ and\ \bibinfo {author}
  {\bibfnamefont {E.}~\bibnamefont {Mamontov}},\ }\href@noop {} {\bibfield
  {journal} {\bibinfo  {journal} {Proc. Natl. Acad. Sci. USA}\ }\textbf
  {\bibinfo {volume} {103}},\ \bibinfo {pages} {9012} (\bibinfo {year}
  {2006})}\BibitemShut {NoStop}%
\bibitem [{\citenamefont {Yoshida}\ \emph {et~al.}(2008)\citenamefont
  {Yoshida}, \citenamefont {Yamaguchi}, \citenamefont {Kittaka}, \citenamefont
  {Bellissent-Funel},\ and\ \citenamefont {Fouquet}}]{Yoshida08}%
  \BibitemOpen
  \bibfield  {author} {\bibinfo {author} {\bibfnamefont {K.}~\bibnamefont
  {Yoshida}}, \bibinfo {author} {\bibfnamefont {T.}~\bibnamefont {Yamaguchi}},
  \bibinfo {author} {\bibfnamefont {S.}~\bibnamefont {Kittaka}}, \bibinfo
  {author} {\bibfnamefont {M.-C.}\ \bibnamefont {Bellissent-Funel}}, \ and\
  \bibinfo {author} {\bibfnamefont {P.}~\bibnamefont {Fouquet}},\ }\href@noop
  {} {\bibfield  {journal} {\bibinfo  {journal} {J. Chem. Phys.}\ }\textbf
  {\bibinfo {volume} {129}},\ \bibinfo {pages} {054702} (\bibinfo {year}
  {2008})}\BibitemShut {NoStop}%
\bibitem [{\citenamefont {Kumar}\ \emph {et~al.}(2006)\citenamefont {Kumar},
  \citenamefont {Yan}, \citenamefont {Xu}, \citenamefont {Mazza}, \citenamefont
  {Buldyrev}, \citenamefont {Chen}, \citenamefont {Sastry},\ and\ \citenamefont
  {Stanley}}]{Kumar06}%
  \BibitemOpen
  \bibfield  {author} {\bibinfo {author} {\bibfnamefont {P.}~\bibnamefont
  {Kumar}}, \bibinfo {author} {\bibfnamefont {Z.}~\bibnamefont {Yan}}, \bibinfo
  {author} {\bibfnamefont {L.}~\bibnamefont {Xu}}, \bibinfo {author}
  {\bibfnamefont {M.~G.}\ \bibnamefont {Mazza}}, \bibinfo {author}
  {\bibfnamefont {S.~V.}\ \bibnamefont {Buldyrev}}, \bibinfo {author}
  {\bibfnamefont {S.-H.}\ \bibnamefont {Chen}}, \bibinfo {author}
  {\bibfnamefont {S.}~\bibnamefont {Sastry}}, \ and\ \bibinfo {author}
  {\bibfnamefont {H.~E.}\ \bibnamefont {Stanley}},\ }\href@noop {} {\bibfield
  {journal} {\bibinfo  {journal} {Phys. Rev. Lett.}\ }\textbf {\bibinfo
  {volume} {97}},\ \bibinfo {pages} {177802} (\bibinfo {year}
  {2006})}\BibitemShut {NoStop}%
\bibitem [{\citenamefont {Gallo}\ \emph {et~al.}(2010)\citenamefont {Gallo},
  \citenamefont {Rovere},\ and\ \citenamefont {Chen}}]{Gallo10}%
  \BibitemOpen
  \bibfield  {author} {\bibinfo {author} {\bibfnamefont {P.}~\bibnamefont
  {Gallo}}, \bibinfo {author} {\bibfnamefont {M.}~\bibnamefont {Rovere}}, \
  and\ \bibinfo {author} {\bibfnamefont {S.-H.}\ \bibnamefont {Chen}},\
  }\href@noop {} {\bibfield  {journal} {\bibinfo  {journal} {J. Phys. Chem.
  Lett.}\ }\textbf {\bibinfo {volume} {1}},\ \bibinfo {pages} {729} (\bibinfo
  {year} {2010})}\BibitemShut {NoStop}%
\bibitem [{\citenamefont {Cerveny}\ \emph {et~al.}(2004)\citenamefont
  {Cerveny}, \citenamefont {Schwartz}, \citenamefont {Bergman},\ and\
  \citenamefont {Swenson}}]{Cerveny04}%
  \BibitemOpen
  \bibfield  {author} {\bibinfo {author} {\bibfnamefont {S.}~\bibnamefont
  {Cerveny}}, \bibinfo {author} {\bibfnamefont {G.~A.}\ \bibnamefont
  {Schwartz}}, \bibinfo {author} {\bibfnamefont {R.}~\bibnamefont {Bergman}}, \
  and\ \bibinfo {author} {\bibfnamefont {J.}~\bibnamefont {Swenson}},\
  }\href@noop {} {\bibfield  {journal} {\bibinfo  {journal} {Phys. Rev. Lett.}\
  }\textbf {\bibinfo {volume} {93}},\ \bibinfo {pages} {245702} (\bibinfo
  {year} {2004})}\BibitemShut {NoStop}%
\bibitem [{\citenamefont {Swenson}\ \emph {et~al.}(2006)\citenamefont
  {Swenson}, \citenamefont {Jansson},\ and\ \citenamefont
  {Bergman}}]{Swenson06}%
  \BibitemOpen
  \bibfield  {author} {\bibinfo {author} {\bibfnamefont {J.}~\bibnamefont
  {Swenson}}, \bibinfo {author} {\bibfnamefont {H.}~\bibnamefont {Jansson}}, \
  and\ \bibinfo {author} {\bibfnamefont {R.}~\bibnamefont {Bergman}},\
  }\href@noop {} {\bibfield  {journal} {\bibinfo  {journal} {Phys. Rev. Lett.}\
  }\textbf {\bibinfo {volume} {96}},\ \bibinfo {pages} {247802} (\bibinfo
  {year} {2006})}\BibitemShut {NoStop}%
\bibitem [{\citenamefont {Hedstr\"om}\ \emph {et~al.}(2007)\citenamefont
  {Hedstr\"om}, \citenamefont {Swenson}, \citenamefont {Bergman}, \citenamefont
  {Jansson},\ and\ \citenamefont {Kittaka}}]{Hedstrom07}%
  \BibitemOpen
  \bibfield  {author} {\bibinfo {author} {\bibfnamefont {J.}~\bibnamefont
  {Hedstr\"om}}, \bibinfo {author} {\bibfnamefont {J.}~\bibnamefont {Swenson}},
  \bibinfo {author} {\bibfnamefont {R.}~\bibnamefont {Bergman}}, \bibinfo
  {author} {\bibfnamefont {H.}~\bibnamefont {Jansson}}, \ and\ \bibinfo
  {author} {\bibfnamefont {S.}~\bibnamefont {Kittaka}},\ }\href@noop {}
  {\bibfield  {journal} {\bibinfo  {journal} {Eur. Phys. J. Special Topics}\
  }\textbf {\bibinfo {volume} {141}},\ \bibinfo {pages} {53} (\bibinfo {year}
  {2007})}\BibitemShut {NoStop}%
\bibitem [{\citenamefont {Cerveny}\ \emph {et~al.}(2008)\citenamefont
  {Cerveny}, \citenamefont {Alegria},\ and\ \citenamefont
  {Colmenero}}]{Cerveny08}%
  \BibitemOpen
  \bibfield  {author} {\bibinfo {author} {\bibfnamefont {S.}~\bibnamefont
  {Cerveny}}, \bibinfo {author} {\bibfnamefont {A.}~\bibnamefont {Alegria}}, \
  and\ \bibinfo {author} {\bibfnamefont {J.}~\bibnamefont {Colmenero}},\
  }\href@noop {} {\bibfield  {journal} {\bibinfo  {journal} {Phys. Rev. E}\
  }\textbf {\bibinfo {volume} {77}},\ \bibinfo {pages} {031803} (\bibinfo
  {year} {2008})}\BibitemShut {NoStop}%
\bibitem [{\citenamefont {Khodadadi}\ \emph {et~al.}(2008)\citenamefont
  {Khodadadi}, \citenamefont {Pawlus},\ and\ \citenamefont
  {Sokolov}}]{Khodadadi08}%
  \BibitemOpen
  \bibfield  {author} {\bibinfo {author} {\bibfnamefont {S.}~\bibnamefont
  {Khodadadi}}, \bibinfo {author} {\bibfnamefont {S.}~\bibnamefont {Pawlus}}, \
  and\ \bibinfo {author} {\bibfnamefont {A.~P.}\ \bibnamefont {Sokolov}},\
  }\href@noop {} {\bibfield  {journal} {\bibinfo  {journal} {J. Phys. Chem. B}\
  }\textbf {\bibinfo {volume} {112}},\ \bibinfo {pages} {14273} (\bibinfo
  {year} {2008})}\BibitemShut {NoStop}%
\bibitem [{\citenamefont {Pawlus}\ \emph {et~al.}(2008)\citenamefont {Pawlus},
  \citenamefont {Khodadadi},\ and\ \citenamefont {Sokolov}}]{Pawlus08}%
  \BibitemOpen
  \bibfield  {author} {\bibinfo {author} {\bibfnamefont {S.}~\bibnamefont
  {Pawlus}}, \bibinfo {author} {\bibfnamefont {S.}~\bibnamefont {Khodadadi}}, \
  and\ \bibinfo {author} {\bibfnamefont {A.~P.}\ \bibnamefont {Sokolov}},\
  }\href@noop {} {\bibfield  {journal} {\bibinfo  {journal} {Phys. Rev. Lett.}\
  }\textbf {\bibinfo {volume} {100}},\ \bibinfo {pages} {108103} (\bibinfo
  {year} {2008})}\BibitemShut {NoStop}%
\bibitem [{\citenamefont {Bruni}\ \emph {et~al.}(2011)\citenamefont {Bruni},
  \citenamefont {Mancinelli},\ and\ \citenamefont {Ricci}}]{Bruni11}%
  \BibitemOpen
  \bibfield  {author} {\bibinfo {author} {\bibfnamefont {F.}~\bibnamefont
  {Bruni}}, \bibinfo {author} {\bibfnamefont {R.}~\bibnamefont {Mancinelli}}, \
  and\ \bibinfo {author} {\bibfnamefont {M.~A.}\ \bibnamefont {Ricci}},\
  }\href@noop {} {\bibfield  {journal} {\bibinfo  {journal} {Phys. Chem. Chem.
  Phys.}\ }\textbf {\bibinfo {volume} {13}},\ \bibinfo {pages} {19773}
  (\bibinfo {year} {2011})}\BibitemShut {NoStop}%
\bibitem [{\citenamefont {Gainaru}\ \emph {et~al.}(2009)\citenamefont
  {Gainaru}, \citenamefont {Fillmer},\ and\ \citenamefont
  {B\"ohmer}}]{Gainaru09}%
  \BibitemOpen
  \bibfield  {author} {\bibinfo {author} {\bibfnamefont {C.}~\bibnamefont
  {Gainaru}}, \bibinfo {author} {\bibfnamefont {A.}~\bibnamefont {Fillmer}}, \
  and\ \bibinfo {author} {\bibfnamefont {R.}~\bibnamefont {B\"ohmer}},\
  }\href@noop {} {\bibfield  {journal} {\bibinfo  {journal} {J. Phys. Chem.
  Lett.}\ }\textbf {\bibinfo {volume} {113}},\ \bibinfo {pages} {12628}
  (\bibinfo {year} {2009})}\BibitemShut {NoStop}%
\bibitem [{\citenamefont {Lusceac}\ \emph {et~al.}(2011)\citenamefont
  {Lusceac}, \citenamefont {Rosenstihl}, \citenamefont {Vogel}, \citenamefont
  {Gainaru}, \citenamefont {Fillmer},\ and\ \citenamefont
  {B\"ohmer}}]{Lusceac11}%
  \BibitemOpen
  \bibfield  {author} {\bibinfo {author} {\bibfnamefont {S.~A.}\ \bibnamefont
  {Lusceac}}, \bibinfo {author} {\bibfnamefont {M.}~\bibnamefont {Rosenstihl}},
  \bibinfo {author} {\bibfnamefont {M.}~\bibnamefont {Vogel}}, \bibinfo
  {author} {\bibfnamefont {C.}~\bibnamefont {Gainaru}}, \bibinfo {author}
  {\bibfnamefont {A.}~\bibnamefont {Fillmer}}, \ and\ \bibinfo {author}
  {\bibfnamefont {R.}~\bibnamefont {B\"ohmer}},\ }\href@noop {} {\bibfield
  {journal} {\bibinfo  {journal} {J. Non-Cryst. Solids}\ }\textbf {\bibinfo
  {volume} {357}},\ \bibinfo {pages} {655} (\bibinfo {year}
  {2011})}\BibitemShut {NoStop}%
\bibitem [{\citenamefont {Vogel}(2008)}]{Vogel08}%
  \BibitemOpen
  \bibfield  {author} {\bibinfo {author} {\bibfnamefont {M.}~\bibnamefont
  {Vogel}},\ }\href@noop {} {\bibfield  {journal} {\bibinfo  {journal} {Phys.
  Rev. Lett.}\ }\textbf {\bibinfo {volume} {101}},\ \bibinfo {pages} {225701}
  (\bibinfo {year} {2008})}\BibitemShut {NoStop}%
\bibitem [{\citenamefont {Sattig}\ and\ \citenamefont
  {Vogel}(2014)}]{Sattig14}%
  \BibitemOpen
  \bibfield  {author} {\bibinfo {author} {\bibfnamefont {M.}~\bibnamefont
  {Sattig}}\ and\ \bibinfo {author} {\bibfnamefont {M.}~\bibnamefont {Vogel}},\
  }\href@noop {} {\bibfield  {journal} {\bibinfo  {journal} {J. Phys. Chem.
  Lett.}\ }\textbf {\bibinfo {volume} {5}},\ \bibinfo {pages} {174} (\bibinfo
  {year} {2014})}\BibitemShut {NoStop}%
\bibitem [{\citenamefont {Doster}\ \emph {et~al.}(2010)\citenamefont {Doster},
  \citenamefont {Busch}, \citenamefont {Gaspar}, \citenamefont {Appavou},
  \citenamefont {Wuttke},\ and\ \citenamefont {Scheer}}]{Doster10}%
  \BibitemOpen
  \bibfield  {author} {\bibinfo {author} {\bibfnamefont {W.}~\bibnamefont
  {Doster}}, \bibinfo {author} {\bibfnamefont {S.}~\bibnamefont {Busch}},
  \bibinfo {author} {\bibfnamefont {A.~M.}\ \bibnamefont {Gaspar}}, \bibinfo
  {author} {\bibfnamefont {M.-S.}\ \bibnamefont {Appavou}}, \bibinfo {author}
  {\bibfnamefont {J.}~\bibnamefont {Wuttke}}, \ and\ \bibinfo {author}
  {\bibfnamefont {H.}~\bibnamefont {Scheer}},\ }\href@noop {} {\bibfield
  {journal} {\bibinfo  {journal} {Phys. Rev. Lett.}\ }\textbf {\bibinfo
  {volume} {104}},\ \bibinfo {pages} {098101} (\bibinfo {year}
  {2010})}\BibitemShut {NoStop}%
\bibitem [{\citenamefont {Ngai}\ \emph {et~al.}(2010)\citenamefont {Ngai},
  \citenamefont {Capaccioli}, \citenamefont {Thayyil},\ and\ \citenamefont
  {Shinashiki}}]{Ngai10}%
  \BibitemOpen
  \bibfield  {author} {\bibinfo {author} {\bibfnamefont {K.~L.}\ \bibnamefont
  {Ngai}}, \bibinfo {author} {\bibfnamefont {S.}~\bibnamefont {Capaccioli}},
  \bibinfo {author} {\bibfnamefont {M.~S.}\ \bibnamefont {Thayyil}}, \ and\
  \bibinfo {author} {\bibfnamefont {N.}~\bibnamefont {Shinashiki}},\
  }\href@noop {} {\bibfield  {journal} {\bibinfo  {journal} {J. Therm. Anal.
  Calorim.}\ }\textbf {\bibinfo {volume} {99}},\ \bibinfo {pages} {123}
  (\bibinfo {year} {2010})}\BibitemShut {NoStop}%
\bibitem [{\citenamefont {Capaccioli}\ \emph {et~al.}(2011)\citenamefont
  {Capaccioli}, \citenamefont {Ngai}, \citenamefont {Ancherbak}, \citenamefont
  {Rolla},\ and\ \citenamefont {N.Shinyashiki}}]{Capaccioli11}%
  \BibitemOpen
  \bibfield  {author} {\bibinfo {author} {\bibfnamefont {S.}~\bibnamefont
  {Capaccioli}}, \bibinfo {author} {\bibfnamefont {K.}~\bibnamefont {Ngai}},
  \bibinfo {author} {\bibfnamefont {S.}~\bibnamefont {Ancherbak}}, \bibinfo
  {author} {\bibfnamefont {P.}~\bibnamefont {Rolla}}, \ and\ \bibinfo {author}
  {\bibnamefont {N.Shinyashiki}},\ }\href@noop {} {\bibfield  {journal}
  {\bibinfo  {journal} {J. Non-Cryst. Solids}\ }\textbf {\bibinfo {volume}
  {357}},\ \bibinfo {pages} {641} (\bibinfo {year} {2011})}\BibitemShut
  {NoStop}%
\bibitem [{\citenamefont {Limmer}\ and\ \citenamefont
  {Chandler}(2011)}]{Limmer11}%
  \BibitemOpen
  \bibfield  {author} {\bibinfo {author} {\bibfnamefont {D.~T.}\ \bibnamefont
  {Limmer}}\ and\ \bibinfo {author} {\bibfnamefont {D.}~\bibnamefont
  {Chandler}},\ }\href@noop {} {\bibfield  {journal} {\bibinfo  {journal} {J.
  Chem. Phys.}\ }\textbf {\bibinfo {volume} {135}},\ \bibinfo {pages} {134503}
  (\bibinfo {year} {2011})}\BibitemShut {NoStop}%
\bibitem [{\citenamefont {Mallamace}\ \emph {et~al.}(2008)\citenamefont
  {Mallamace}, \citenamefont {Corsaro}, \citenamefont {Broccio}, \citenamefont
  {Branca}, \citenamefont {Gonzalez-Segredo}, \citenamefont {Spooren},
  \citenamefont {Chen},\ and\ \citenamefont {Stanley}}]{Mallamace08}%
  \BibitemOpen
  \bibfield  {author} {\bibinfo {author} {\bibfnamefont {F.}~\bibnamefont
  {Mallamace}}, \bibinfo {author} {\bibfnamefont {C.}~\bibnamefont {Corsaro}},
  \bibinfo {author} {\bibfnamefont {M.}~\bibnamefont {Broccio}}, \bibinfo
  {author} {\bibfnamefont {C.}~\bibnamefont {Branca}}, \bibinfo {author}
  {\bibfnamefont {N.}~\bibnamefont {Gonzalez-Segredo}}, \bibinfo {author}
  {\bibfnamefont {J.}~\bibnamefont {Spooren}}, \bibinfo {author} {\bibfnamefont
  {S.-H.}\ \bibnamefont {Chen}}, \ and\ \bibinfo {author} {\bibfnamefont
  {H.~E.}\ \bibnamefont {Stanley}},\ }\href@noop {} {\bibfield  {journal}
  {\bibinfo  {journal} {Proc. Natl. Acad. Sci. USA}\ }\textbf {\bibinfo
  {volume} {105}},\ \bibinfo {pages} {12725} (\bibinfo {year}
  {2008})}\BibitemShut {NoStop}%
\bibitem [{\citenamefont {Mallamace}\ \emph {et~al.}(2006)\citenamefont
  {Mallamace}, \citenamefont {Broccio}, \citenamefont {Corsaro}, \citenamefont
  {Faraone}, \citenamefont {Wanderlingh}, \citenamefont {Liu}, \citenamefont
  {Mou},\ and\ \citenamefont {S.-H.Chen}}]{Mallamace06}%
  \BibitemOpen
  \bibfield  {author} {\bibinfo {author} {\bibfnamefont {F.}~\bibnamefont
  {Mallamace}}, \bibinfo {author} {\bibfnamefont {M.}~\bibnamefont {Broccio}},
  \bibinfo {author} {\bibfnamefont {C.}~\bibnamefont {Corsaro}}, \bibinfo
  {author} {\bibfnamefont {A.}~\bibnamefont {Faraone}}, \bibinfo {author}
  {\bibfnamefont {U.}~\bibnamefont {Wanderlingh}}, \bibinfo {author}
  {\bibfnamefont {L.}~\bibnamefont {Liu}}, \bibinfo {author} {\bibfnamefont
  {C.-Y.}\ \bibnamefont {Mou}}, \ and\ \bibinfo {author} {\bibnamefont
  {S.-H.Chen}},\ }\href@noop {} {\bibfield  {journal} {\bibinfo  {journal} {J.
  Chem. Phys.}\ }\textbf {\bibinfo {volume} {124}},\ \bibinfo {pages} {161102}
  (\bibinfo {year} {2006})}\BibitemShut {NoStop}%
\bibitem [{\citenamefont {Mallamace}\ \emph {et~al.}(2007)\citenamefont
  {Mallamace}, \citenamefont {S.-H.Chen}, \citenamefont {Broccio},
  \citenamefont {Corsaro}, \citenamefont {Crupi}, \citenamefont {Majolino}, ,
  \citenamefont {Venuti}, \citenamefont {Baglioni}, \citenamefont {Fratini},
  \citenamefont {Vannucci},\ and\ \citenamefont {Stanley}}]{Mallamace07}%
  \BibitemOpen
  \bibfield  {author} {\bibinfo {author} {\bibfnamefont {F.}~\bibnamefont
  {Mallamace}}, \bibinfo {author} {\bibnamefont {S.-H.Chen}}, \bibinfo {author}
  {\bibfnamefont {M.}~\bibnamefont {Broccio}}, \bibinfo {author} {\bibfnamefont
  {C.}~\bibnamefont {Corsaro}}, \bibinfo {author} {\bibfnamefont
  {V.}~\bibnamefont {Crupi}}, \bibinfo {author} {\bibfnamefont
  {D.}~\bibnamefont {Majolino}}, , \bibinfo {author} {\bibfnamefont
  {V.}~\bibnamefont {Venuti}}, \bibinfo {author} {\bibfnamefont
  {P.}~\bibnamefont {Baglioni}}, \bibinfo {author} {\bibfnamefont
  {E.}~\bibnamefont {Fratini}}, \bibinfo {author} {\bibfnamefont
  {C.}~\bibnamefont {Vannucci}}, \ and\ \bibinfo {author} {\bibfnamefont
  {H.~E.}\ \bibnamefont {Stanley}},\ }\href@noop {} {\bibfield  {journal}
  {\bibinfo  {journal} {J. Chem. Phys.}\ }\textbf {\bibinfo {volume} {127}},\
  \bibinfo {pages} {045104} (\bibinfo {year} {2007})}\BibitemShut {NoStop}%
\bibitem [{\citenamefont {Rosenstihl}\ and\ \citenamefont
  {Vogel}(2011)}]{Rosenstihl11}%
  \BibitemOpen
  \bibfield  {author} {\bibinfo {author} {\bibfnamefont {M.}~\bibnamefont
  {Rosenstihl}}\ and\ \bibinfo {author} {\bibfnamefont {M.}~\bibnamefont
  {Vogel}},\ }\href@noop {} {\bibfield  {journal} {\bibinfo  {journal} {J.
  Chem. Phys.}\ }\textbf {\bibinfo {volume} {135}},\ \bibinfo {pages} {164503}
  (\bibinfo {year} {2011})}\BibitemShut {NoStop}%
\bibitem [{\citenamefont {Hwang}\ \emph {et~al.}(2007)\citenamefont {Hwang},
  \citenamefont {Chu}, \citenamefont {Sinha},\ and\ \citenamefont
  {Hwang}}]{Hwang07}%
  \BibitemOpen
  \bibfield  {author} {\bibinfo {author} {\bibfnamefont {D.~W.}\ \bibnamefont
  {Hwang}}, \bibinfo {author} {\bibfnamefont {C.-C.}\ \bibnamefont {Chu}},
  \bibinfo {author} {\bibfnamefont {A.~K.}\ \bibnamefont {Sinha}}, \ and\
  \bibinfo {author} {\bibfnamefont {L.-P.}\ \bibnamefont {Hwang}},\ }\href@noop
  {} {\bibfield  {journal} {\bibinfo  {journal} {J. Chem. Phys.}\ }\textbf
  {\bibinfo {volume} {126}},\ \bibinfo {pages} {044702} (\bibinfo {year}
  {2007})}\BibitemShut {NoStop}%
\bibitem [{\citenamefont {Schmidt-Rohr}\ and\ \citenamefont
  {Spie\ss}(1994)}]{Schmidt-Rohr94}%
  \BibitemOpen
  \bibfield  {author} {\bibinfo {author} {\bibfnamefont {K.}~\bibnamefont
  {Schmidt-Rohr}}\ and\ \bibinfo {author} {\bibfnamefont {W.}~\bibnamefont
  {Spie\ss}},\ }\href@noop {} {\emph {\bibinfo {title} {Multidimensional
  Solid-State NMR and Polymers}}}\ (\bibinfo  {publisher} {Academic Press Ltd.,
  London},\ \bibinfo {year} {1994})\BibitemShut {NoStop}%
\bibitem [{\citenamefont {Fleischer}\ and\ \citenamefont
  {Fujara}(1994)}]{Fleischer94}%
  \BibitemOpen
  \bibfield  {author} {\bibinfo {author} {\bibfnamefont {G.}~\bibnamefont
  {Fleischer}}\ and\ \bibinfo {author} {\bibfnamefont {F.}~\bibnamefont
  {Fujara}},\ }in\ \href {\doibase 10.1007/978-3-642-78483-5_4} {\emph
  {\bibinfo {booktitle} {NMR, Basic Principles and Progress}}},\ Vol.~\bibinfo
  {volume} {30},\ \bibinfo {editor} {edited by\ \bibinfo {editor}
  {\bibfnamefont {P.}~\bibnamefont {Diehl}}, \bibinfo {editor} {\bibfnamefont
  {E.}~\bibnamefont {Fluck}}, \bibinfo {editor} {\bibfnamefont
  {H.}~\bibnamefont {G\"unther}}, \bibinfo {editor} {\bibfnamefont
  {R.}~\bibnamefont {Kosfeld}}, \ and\ \bibinfo {editor} {\bibfnamefont
  {J.}~\bibnamefont {Seelig}}}\ (\bibinfo  {publisher} {Springer Berlin
  Heidelberg},\ \bibinfo {year} {1994})\ pp.\ \bibinfo {pages} {159--207},\
  \bibinfo {edition} {1st}\ ed.\BibitemShut {Stop}%
\bibitem [{\citenamefont {B{\"ohmer}}\ \emph {et~al.}(2001)\citenamefont
  {B{\"ohmer}}, \citenamefont {Diezemann}, \citenamefont {Hinze},\ and\
  \citenamefont {R{\"o}ssler}}]{Boehmer01}%
  \BibitemOpen
  \bibfield  {author} {\bibinfo {author} {\bibfnamefont {R.}~\bibnamefont
  {B{\"ohmer}}}, \bibinfo {author} {\bibfnamefont {G.}~\bibnamefont
  {Diezemann}}, \bibinfo {author} {\bibfnamefont {G.}~\bibnamefont {Hinze}}, \
  and\ \bibinfo {author} {\bibfnamefont {E.}~\bibnamefont {R{\"o}ssler}},\
  }\href@noop {} {\bibfield  {journal} {\bibinfo  {journal} {Prog. Nucl. Magn.
  Res. Spectr.}\ }\textbf {\bibinfo {volume} {39}},\ \bibinfo {pages} {191}
  (\bibinfo {year} {2001})}\BibitemShut {NoStop}%
\bibitem [{\citenamefont {Fujara}\ \emph {et~al.}(1986)\citenamefont {Fujara},
  \citenamefont {Wefing},\ and\ \citenamefont {Spiess}}]{Fujara86}%
  \BibitemOpen
  \bibfield  {author} {\bibinfo {author} {\bibfnamefont {F.}~\bibnamefont
  {Fujara}}, \bibinfo {author} {\bibfnamefont {S.}~\bibnamefont {Wefing}}, \
  and\ \bibinfo {author} {\bibfnamefont {H.~W.}\ \bibnamefont {Spiess}},\
  }\href@noop {} {\bibfield  {journal} {\bibinfo  {journal} {J. Chem. Phys.}\
  }\textbf {\bibinfo {volume} {84}},\ \bibinfo {pages} {4579} (\bibinfo {year}
  {1986})}\BibitemShut {NoStop}%
\bibitem [{\citenamefont {Callaghan}(1991)}]{Callaghan91}%
  \BibitemOpen
  \bibfield  {author} {\bibinfo {author} {\bibfnamefont {P.~T.}\ \bibnamefont
  {Callaghan}},\ }\href@noop {} {\emph {\bibinfo {title} {Principles of Nuclear
  Magnetic Resonance Microscopy}}}\ (\bibinfo  {publisher} {Clarendon Press,
  Oxford},\ \bibinfo {year} {1991})\BibitemShut {NoStop}%
\bibitem [{\citenamefont {Klameth}\ and\ \citenamefont
  {Vogel}(2013)}]{Klameth13}%
  \BibitemOpen
  \bibfield  {author} {\bibinfo {author} {\bibfnamefont {F.}~\bibnamefont
  {Klameth}}\ and\ \bibinfo {author} {\bibfnamefont {M.}~\bibnamefont
  {Vogel}},\ }\href@noop {} {\bibfield  {journal} {\bibinfo  {journal} {J.
  Chem. Phys.}\ }\textbf {\bibinfo {volume} {138}},\ \bibinfo {pages} {134503}
  (\bibinfo {year} {2013})}\BibitemShut {NoStop}%
\bibitem [{\citenamefont {Klameth}\ \emph {et~al.}(2014)\citenamefont
  {Klameth}, \citenamefont {Henritzi},\ and\ \citenamefont
  {Vogel}}]{Klameth14}%
  \BibitemOpen
  \bibfield  {author} {\bibinfo {author} {\bibfnamefont {F.}~\bibnamefont
  {Klameth}}, \bibinfo {author} {\bibfnamefont {P.}~\bibnamefont {Henritzi}}, \
  and\ \bibinfo {author} {\bibfnamefont {M.}~\bibnamefont {Vogel}},\
  }\href@noop {} {\bibfield  {journal} {\bibinfo  {journal} {J. Chem. Phys.}\
  }\textbf {\bibinfo {volume} {140}},\ \bibinfo {pages} {144501} (\bibinfo
  {year} {2014})}\BibitemShut {NoStop}%
\bibitem [{\citenamefont {Smolin}\ \emph {et~al.}(2005)\citenamefont {Smolin},
  \citenamefont {Oleinikova}, \citenamefont {Brovchenko}, \citenamefont
  {Geiger},\ and\ \citenamefont {Winter}}]{Smolin05}%
  \BibitemOpen
  \bibfield  {author} {\bibinfo {author} {\bibfnamefont {N.}~\bibnamefont
  {Smolin}}, \bibinfo {author} {\bibfnamefont {A.}~\bibnamefont {Oleinikova}},
  \bibinfo {author} {\bibfnamefont {I.}~\bibnamefont {Brovchenko}}, \bibinfo
  {author} {\bibfnamefont {A.}~\bibnamefont {Geiger}}, \ and\ \bibinfo {author}
  {\bibfnamefont {R.}~\bibnamefont {Winter}},\ }\href@noop {} {\bibfield
  {journal} {\bibinfo  {journal} {J. Phys. Chem. B}\ }\textbf {\bibinfo
  {volume} {109}},\ \bibinfo {pages} {10995} (\bibinfo {year}
  {2005})}\BibitemShut {NoStop}%
\bibitem [{\citenamefont {Berendsen}\ \emph {et~al.}(1981)\citenamefont
  {Berendsen}, \citenamefont {Postma}, \citenamefont {van Gunsteren},\ and\
  \citenamefont {Hermans}}]{SPC}%
  \BibitemOpen
  \bibfield  {author} {\bibinfo {author} {\bibfnamefont {H.}~\bibnamefont
  {Berendsen}}, \bibinfo {author} {\bibfnamefont {J.}~\bibnamefont {Postma}},
  \bibinfo {author} {\bibfnamefont {W.}~\bibnamefont {van Gunsteren}}, \ and\
  \bibinfo {author} {\bibfnamefont {J.}~\bibnamefont {Hermans}},\ }in\
  \href@noop {} {\emph {\bibinfo {booktitle} {Intermolecular Forces}}},\
  \bibinfo {editor} {edited by\ \bibinfo {editor} {\bibfnamefont
  {B.}~\bibnamefont {Pullmann}}}\ (\bibinfo  {publisher} {Reidel, Dordrecht},\
  \bibinfo {year} {1981})\ p.\ \bibinfo {pages} {331}\BibitemShut {NoStop}%
\bibitem [{\citenamefont {Vogel}(2009)}]{Vogel09}%
  \BibitemOpen
  \bibfield  {author} {\bibinfo {author} {\bibfnamefont {M.}~\bibnamefont
  {Vogel}},\ }\href@noop {} {\bibfield  {journal} {\bibinfo  {journal} {J.
  Phys. Chem. B}\ }\textbf {\bibinfo {volume} {113}},\ \bibinfo {pages} {2009}
  (\bibinfo {year} {2009})}\BibitemShut {NoStop}%
\bibitem [{\citenamefont {Berendsen}\ \emph {et~al.}(1987)\citenamefont
  {Berendsen}, \citenamefont {Grigera},\ and\ \citenamefont
  {Straatsma}}]{SPCE}%
  \BibitemOpen
  \bibfield  {author} {\bibinfo {author} {\bibfnamefont {H.~J.~C.}\
  \bibnamefont {Berendsen}}, \bibinfo {author} {\bibfnamefont {J.~R.}\
  \bibnamefont {Grigera}}, \ and\ \bibinfo {author} {\bibfnamefont {T.~P.}\
  \bibnamefont {Straatsma}},\ }\href@noop {} {\bibfield  {journal} {\bibinfo
  {journal} {J.Phys. Chem.}\ }\textbf {\bibinfo {volume} {91}},\ \bibinfo
  {pages} {6269} (\bibinfo {year} {1987})}\BibitemShut {NoStop}%
\bibitem [{\citenamefont {Scheidler}\ \emph {et~al.}(2004)\citenamefont
  {Scheidler}, \citenamefont {Kob},\ and\ \citenamefont
  {Binder}}]{Scheidler04}%
  \BibitemOpen
  \bibfield  {author} {\bibinfo {author} {\bibfnamefont {P.}~\bibnamefont
  {Scheidler}}, \bibinfo {author} {\bibfnamefont {W.}~\bibnamefont {Kob}}, \
  and\ \bibinfo {author} {\bibfnamefont {K.}~\bibnamefont {Binder}},\
  }\href@noop {} {\bibfield  {journal} {\bibinfo  {journal} {J. Phys. Chem. B}\
  }\textbf {\bibinfo {volume} {108}},\ \bibinfo {pages} {6673} (\bibinfo {year}
  {2004})}\BibitemShut {NoStop}%
\bibitem [{\citenamefont {Berthier}\ and\ \citenamefont
  {Kob}(2012)}]{Berthier12}%
  \BibitemOpen
  \bibfield  {author} {\bibinfo {author} {\bibfnamefont {L.}~\bibnamefont
  {Berthier}}\ and\ \bibinfo {author} {\bibfnamefont {W.}~\bibnamefont {Kob}},\
  }\href@noop {} {\bibfield  {journal} {\bibinfo  {journal} {Phys. Rev. E}\
  }\textbf {\bibinfo {volume} {85}},\ \bibinfo {pages} {011102} (\bibinfo
  {year} {2012})}\BibitemShut {NoStop}%
\bibitem [{\citenamefont {Harrach}\ and\ \citenamefont
  {Drossel}(2014)}]{Drossel14}%
  \BibitemOpen
  \bibfield  {author} {\bibinfo {author} {\bibfnamefont {M.~F.}\ \bibnamefont
  {Harrach}}\ and\ \bibinfo {author} {\bibfnamefont {B.}~\bibnamefont
  {Drossel}},\ }\href@noop {} {\bibfield  {journal} {\bibinfo  {journal} {J.
  Chem. Phys.}\ }\textbf {\bibinfo {volume} {140}},\ \bibinfo {pages} {174501}
  (\bibinfo {year} {2014})}\BibitemShut {NoStop}%
\bibitem [{\citenamefont {Lusceac}\ \emph {et~al.}(2010)\citenamefont
  {Lusceac}, \citenamefont {Vogel},\ and\ \citenamefont
  {Herbers}}]{Lusceac10BBA}%
  \BibitemOpen
  \bibfield  {author} {\bibinfo {author} {\bibfnamefont {S.~A.}\ \bibnamefont
  {Lusceac}}, \bibinfo {author} {\bibfnamefont {M.~R.}\ \bibnamefont {Vogel}},
  \ and\ \bibinfo {author} {\bibfnamefont {C.~R.}\ \bibnamefont {Herbers}},\
  }\href@noop {} {\bibfield  {journal} {\bibinfo  {journal} {Biochim. Biophys.
  Acta, Proteins Proteomics}\ }\textbf {\bibinfo {volume} {1804}},\ \bibinfo
  {pages} {41} (\bibinfo {year} {2010})}\BibitemShut {NoStop}%
\bibitem [{\citenamefont {Vogel.}(2010)}]{Vogel10}%
  \BibitemOpen
  \bibfield  {author} {\bibinfo {author} {\bibfnamefont {M.}~\bibnamefont
  {Vogel.}},\ }\href@noop {} {\bibfield  {journal} {\bibinfo  {journal} {Eur.
  Phys. J. Special Topics}\ }\textbf {\bibinfo {volume} {189}},\ \bibinfo
  {pages} {2010} (\bibinfo {year} {2010})}\BibitemShut {NoStop}%
\bibitem [{\citenamefont {Jansson}\ \emph {et~al.}(2011)\citenamefont
  {Jansson}, \citenamefont {Bergman},\ and\ \citenamefont
  {Swenson}}]{Jansson11}%
  \BibitemOpen
  \bibfield  {author} {\bibinfo {author} {\bibfnamefont {H.}~\bibnamefont
  {Jansson}}, \bibinfo {author} {\bibfnamefont {R.}~\bibnamefont {Bergman}}, \
  and\ \bibinfo {author} {\bibfnamefont {J.}~\bibnamefont {Swenson}},\
  }\href@noop {} {\bibfield  {journal} {\bibinfo  {journal} {J. Phys. Chem. B}\
  }\textbf {\bibinfo {volume} {115}},\ \bibinfo {pages} {4099} (\bibinfo {year}
  {2011})}\BibitemShut {NoStop}%
\bibitem [{\citenamefont {Lusceac}\ and\ \citenamefont
  {Vogel}(2010)}]{Lusceac10JPC}%
  \BibitemOpen
  \bibfield  {author} {\bibinfo {author} {\bibfnamefont {S.~A.}\ \bibnamefont
  {Lusceac}}\ and\ \bibinfo {author} {\bibfnamefont {M.~R.}\ \bibnamefont
  {Vogel}},\ }\href@noop {} {\bibfield  {journal} {\bibinfo  {journal} {J.
  Phys. Chem. B}\ }\textbf {\bibinfo {volume} {114}},\ \bibinfo {pages} {10209}
  (\bibinfo {year} {2010})}\BibitemShut {NoStop}%
\bibitem [{\citenamefont {Sjostrom}\ \emph {et~al.}(2008)\citenamefont
  {Sjostrom}, \citenamefont {Swenson}, \citenamefont {Bergman},\ and\
  \citenamefont {Kittaka}}]{Sjostrom08}%
  \BibitemOpen
  \bibfield  {author} {\bibinfo {author} {\bibfnamefont {J.}~\bibnamefont
  {Sjostrom}}, \bibinfo {author} {\bibfnamefont {J.}~\bibnamefont {Swenson}},
  \bibinfo {author} {\bibfnamefont {R.}~\bibnamefont {Bergman}}, \ and\
  \bibinfo {author} {\bibfnamefont {S.}~\bibnamefont {Kittaka}},\ }\href@noop
  {} {\bibfield  {journal} {\bibinfo  {journal} {J. Chem. Phys.}\ }\textbf
  {\bibinfo {volume} {128}},\ \bibinfo {pages} {154503} (\bibinfo {year}
  {2008})}\BibitemShut {NoStop}%
\bibitem [{\citenamefont {W.~Schnauss}\ and\ \citenamefont
  {Sillescu}(1990)}]{Schnauss90}%
  \BibitemOpen
  \bibfield  {author} {\bibinfo {author} {\bibfnamefont {K.~H.}\ \bibnamefont
  {W.~Schnauss}, \bibfnamefont {F.~Fujara}}\ and\ \bibinfo {author}
  {\bibfnamefont {H.}~\bibnamefont {Sillescu}},\ }\href@noop {} {\bibfield
  {journal} {\bibinfo  {journal} {Chem. Phys. Lett.}\ }\textbf {\bibinfo
  {volume} {166}},\ \bibinfo {pages} {381} (\bibinfo {year}
  {1990})}\BibitemShut {NoStop}%
\bibitem [{\citenamefont {Lusceac}\ \emph {et~al.}(2004)\citenamefont
  {Lusceac}, \citenamefont {Koplin}, \citenamefont {Medick}, \citenamefont
  {Vogel}, \citenamefont {Brodie-Linder}, \citenamefont {LeQuellec},
  \citenamefont {Alba-Simionesco},\ and\ \citenamefont {E.~A}}]{Lusceac04}%
  \BibitemOpen
  \bibfield  {author} {\bibinfo {author} {\bibfnamefont {S.~A.}\ \bibnamefont
  {Lusceac}}, \bibinfo {author} {\bibfnamefont {C.}~\bibnamefont {Koplin}},
  \bibinfo {author} {\bibfnamefont {P.}~\bibnamefont {Medick}}, \bibinfo
  {author} {\bibfnamefont {M.}~\bibnamefont {Vogel}}, \bibinfo {author}
  {\bibfnamefont {N.}~\bibnamefont {Brodie-Linder}}, \bibinfo {author}
  {\bibfnamefont {C.}~\bibnamefont {LeQuellec}}, \bibinfo {author}
  {\bibfnamefont {C.}~\bibnamefont {Alba-Simionesco}}, \ and\ \bibinfo {author}
  {\bibfnamefont {R.}~\bibnamefont {E.~A}},\ }\href@noop {} {\bibfield
  {journal} {\bibinfo  {journal} {J. Phys. Chem. B}\ }\textbf {\bibinfo
  {volume} {108}},\ \bibinfo {pages} {16601} (\bibinfo {year}
  {2004})}\BibitemShut {NoStop}%
\bibitem [{\citenamefont {Zorn}(2002)}]{Zorn02}%
  \BibitemOpen
  \bibfield  {author} {\bibinfo {author} {\bibfnamefont {R.}~\bibnamefont
  {Zorn}},\ }\href@noop {} {\bibfield  {journal} {\bibinfo  {journal} {J. Chem.
  Phys.}\ }\textbf {\bibinfo {volume} {116}},\ \bibinfo {pages} {3204}
  (\bibinfo {year} {2002})}\BibitemShut {NoStop}%
\bibitem [{\citenamefont {B{\"ohmer}}\ and\ \citenamefont
  {Hinze}(1998)}]{Boehmer98}%
  \BibitemOpen
  \bibfield  {author} {\bibinfo {author} {\bibfnamefont {R.}~\bibnamefont
  {B{\"ohmer}}}\ and\ \bibinfo {author} {\bibfnamefont {G.}~\bibnamefont
  {Hinze}},\ }\href@noop {} {\bibfield  {journal} {\bibinfo  {journal} {J.
  Chem. Phys.}\ }\textbf {\bibinfo {volume} {109}},\ \bibinfo {pages} {241}
  (\bibinfo {year} {1998})}\BibitemShut {NoStop}%
\bibitem [{\citenamefont {Hinze}(1998)}]{Hinze98}%
  \BibitemOpen
  \bibfield  {author} {\bibinfo {author} {\bibfnamefont {G.}~\bibnamefont
  {Hinze}},\ }\href@noop {} {\bibfield  {journal} {\bibinfo  {journal} {Phys.
  Rev. E}\ }\textbf {\bibinfo {volume} {57}},\ \bibinfo {pages} {2010}
  (\bibinfo {year} {1998})}\BibitemShut {NoStop}%
\bibitem [{\citenamefont {Wachner}\ and\ \citenamefont
  {Jeffrey}(1999)}]{Wachner99}%
  \BibitemOpen
  \bibfield  {author} {\bibinfo {author} {\bibfnamefont {A.~M.}\ \bibnamefont
  {Wachner}}\ and\ \bibinfo {author} {\bibfnamefont {K.~R.}\ \bibnamefont
  {Jeffrey}},\ }\href@noop {} {\bibfield  {journal} {\bibinfo  {journal} {J.
  Chem. Phys.}\ }\textbf {\bibinfo {volume} {111}},\ \bibinfo {pages} {10611}
  (\bibinfo {year} {1999})}\BibitemShut {NoStop}%
\bibitem [{\citenamefont {K\"ampf}\ \emph {et~al.}(2012)\citenamefont
  {K\"ampf}, \citenamefont {Klameth},\ and\ \citenamefont {Vogel}}]{Kampf12}%
  \BibitemOpen
  \bibfield  {author} {\bibinfo {author} {\bibfnamefont {K.}~\bibnamefont
  {K\"ampf}}, \bibinfo {author} {\bibfnamefont {F.}~\bibnamefont {Klameth}}, \
  and\ \bibinfo {author} {\bibfnamefont {M.}~\bibnamefont {Vogel}},\
  }\href@noop {} {\bibfield  {journal} {\bibinfo  {journal} {J. Chem. Phys.}\
  }\textbf {\bibinfo {volume} {137}},\ \bibinfo {pages} {205105} (\bibinfo
  {year} {2012})}\BibitemShut {NoStop}%
\bibitem [{\citenamefont {K\"ampf}\ \emph {et~al.}(2014)\citenamefont
  {K\"ampf}, \citenamefont {Kremmling},\ and\ \citenamefont {Vogel}}]{Kampf14}%
  \BibitemOpen
  \bibfield  {author} {\bibinfo {author} {\bibfnamefont {K.}~\bibnamefont
  {K\"ampf}}, \bibinfo {author} {\bibfnamefont {B.}~\bibnamefont {Kremmling}},
  \ and\ \bibinfo {author} {\bibfnamefont {M.}~\bibnamefont {Vogel}},\
  }\href@noop {} {\bibfield  {journal} {\bibinfo  {journal} {Phys. Rev. E}\
  }\textbf {\bibinfo {volume} {89}},\ \bibinfo {pages} {032710} (\bibinfo
  {year} {2014})}\BibitemShut {NoStop}%
\bibitem [{\citenamefont {Lagi}\ \emph {et~al.}(2009)\citenamefont {Lagi},
  \citenamefont {Baglioni},\ and\ \citenamefont {Chen}}]{Lagi09}%
  \BibitemOpen
  \bibfield  {author} {\bibinfo {author} {\bibfnamefont {M.}~\bibnamefont
  {Lagi}}, \bibinfo {author} {\bibfnamefont {P.}~\bibnamefont {Baglioni}}, \
  and\ \bibinfo {author} {\bibfnamefont {S.-H.}\ \bibnamefont {Chen}},\
  }\href@noop {} {\bibfield  {journal} {\bibinfo  {journal} {Phys. Rev. Lett.}\
  }\textbf {\bibinfo {volume} {103}},\ \bibinfo {pages} {108102} (\bibinfo
  {year} {2009})}\BibitemShut {NoStop}%
\bibitem [{\citenamefont {Kneller}\ and\ \citenamefont
  {Hinsen}(2004)}]{Kneller04}%
  \BibitemOpen
  \bibfield  {author} {\bibinfo {author} {\bibfnamefont {G.~R.}\ \bibnamefont
  {Kneller}}\ and\ \bibinfo {author} {\bibfnamefont {K.}~\bibnamefont
  {Hinsen}},\ }\href@noop {} {\bibfield  {journal} {\bibinfo  {journal} {J.
  Chem. Phys.}\ }\textbf {\bibinfo {volume} {121}},\ \bibinfo {pages} {10278}
  (\bibinfo {year} {2004})}\BibitemShut {NoStop}%
\bibitem [{\citenamefont {Metzler}\ and\ \citenamefont
  {Klafter}(2000)}]{Metzler00}%
  \BibitemOpen
  \bibfield  {author} {\bibinfo {author} {\bibfnamefont {R.}~\bibnamefont
  {Metzler}}\ and\ \bibinfo {author} {\bibfnamefont {J.}~\bibnamefont
  {Klafter}},\ }\href@noop {} {\bibfield  {journal} {\bibinfo  {journal} {Phys.
  Rep.}\ }\textbf {\bibinfo {volume} {339}},\ \bibinfo {pages} {1} (\bibinfo
  {year} {2000})}\BibitemShut {NoStop}%
\bibitem [{\citenamefont {Vogel}\ \emph {et~al.}(2005)\citenamefont {Vogel},
  \citenamefont {Medick},\ and\ \citenamefont {R\"ossler}}]{Vogel05}%
  \BibitemOpen
  \bibfield  {author} {\bibinfo {author} {\bibfnamefont {M.}~\bibnamefont
  {Vogel}}, \bibinfo {author} {\bibfnamefont {P.}~\bibnamefont {Medick}}, \
  and\ \bibinfo {author} {\bibfnamefont {E.}~\bibnamefont {R\"ossler}},\
  }\href@noop {} {\bibfield  {journal} {\bibinfo  {journal} {Ann. Rep. NMR
  Spectrosc.}\ }\textbf {\bibinfo {volume} {56}},\ \bibinfo {pages} {231}
  (\bibinfo {year} {2005})}\BibitemShut {NoStop}%
\bibitem [{\citenamefont {B\"ohmer}\ \emph {et~al.}(2007)\citenamefont
  {B\"ohmer}, \citenamefont {Jeffrey},\ and\ \citenamefont
  {Vogel}}]{Boehmer07}%
  \BibitemOpen
  \bibfield  {author} {\bibinfo {author} {\bibfnamefont {R.}~\bibnamefont
  {B\"ohmer}}, \bibinfo {author} {\bibfnamefont {K.~R.}\ \bibnamefont
  {Jeffrey}}, \ and\ \bibinfo {author} {\bibfnamefont {M.}~\bibnamefont
  {Vogel}},\ }\href@noop {} {\bibfield  {journal} {\bibinfo  {journal} {Prog.
  Nucl. Mag. Res. Sp.}\ }\textbf {\bibinfo {volume} {50}},\ \bibinfo {pages}
  {87} (\bibinfo {year} {2007})}\BibitemShut {NoStop}%
\bibitem [{\citenamefont {Brinkmann}\ \emph {et~al.}(2010)\citenamefont
  {Brinkmann}, \citenamefont {Faske}, \citenamefont {Koch},\ and\ \citenamefont
  {Vogel}}]{Brinkmann10}%
  \BibitemOpen
  \bibfield  {author} {\bibinfo {author} {\bibfnamefont {C.}~\bibnamefont
  {Brinkmann}}, \bibinfo {author} {\bibfnamefont {S.}~\bibnamefont {Faske}},
  \bibinfo {author} {\bibfnamefont {B.}~\bibnamefont {Koch}}, \ and\ \bibinfo
  {author} {\bibfnamefont {M.}~\bibnamefont {Vogel}},\ }\href@noop {}
  {\bibfield  {journal} {\bibinfo  {journal} {Z. Phys. Chem.}\ }\textbf
  {\bibinfo {volume} {224}},\ \bibinfo {pages} {1535} (\bibinfo {year}
  {2010})}\BibitemShut {NoStop}%
\bibitem [{\citenamefont {Sj\"ostr\"om}\ \emph {et~al.}(2010)\citenamefont
  {Sj\"ostr\"om}, \citenamefont {Mattsson}, \citenamefont {Bergman},
  \citenamefont {Johansson}, \citenamefont {Josefsson}, \citenamefont
  {Svantesson},\ and\ \citenamefont {Swenson}}]{Sjostrom10}%
  \BibitemOpen
  \bibfield  {author} {\bibinfo {author} {\bibfnamefont {J.}~\bibnamefont
  {Sj\"ostr\"om}}, \bibinfo {author} {\bibfnamefont {J.}~\bibnamefont
  {Mattsson}}, \bibinfo {author} {\bibfnamefont {R.}~\bibnamefont {Bergman}},
  \bibinfo {author} {\bibfnamefont {E.}~\bibnamefont {Johansson}}, \bibinfo
  {author} {\bibfnamefont {K.}~\bibnamefont {Josefsson}}, \bibinfo {author}
  {\bibfnamefont {D.}~\bibnamefont {Svantesson}}, \ and\ \bibinfo {author}
  {\bibfnamefont {J.}~\bibnamefont {Swenson}},\ }\href@noop {} {\bibfield
  {journal} {\bibinfo  {journal} {Phys. Chem. Chem. Phys.}\ }\textbf {\bibinfo
  {volume} {12}},\ \bibinfo {pages} {10452} (\bibinfo {year}
  {2010})}\BibitemShut {NoStop}%
\bibitem [{\citenamefont {Swenson}\ and\ \citenamefont
  {Teixeira}(2010)}]{Swenson10}%
  \BibitemOpen
  \bibfield  {author} {\bibinfo {author} {\bibfnamefont {J.}~\bibnamefont
  {Swenson}}\ and\ \bibinfo {author} {\bibfnamefont {J.}~\bibnamefont
  {Teixeira}},\ }\href@noop {} {\bibfield  {journal} {\bibinfo  {journal} {J.
  Chem. Phys.}\ }\textbf {\bibinfo {volume} {132}},\ \bibinfo {pages} {104508}
  (\bibinfo {year} {2010})}\BibitemShut {NoStop}%
\end{thebibliography}%

\end{document}